\documentclass{aa}  

\usepackage{graphicx}
\usepackage{tikz}
\usetikzlibrary{shapes, arrows.meta, positioning}
\usepackage{txfonts}
\usepackage{xcolor}
\usepackage{lipsum}
\usepackage{hyperref}
\usepackage{graphicx}
\usepackage{caption}
\usepackage{subcaption}
\usepackage{float}
\usepackage{threeparttable}
\usepackage{stfloats}
\usepackage{xspace}
\newcommand{\GWT}{\texttt{GWToolbox}\xspace}
\newcommand{\Tb}{\texttt{Toolbox}\xspace}

\begin{document}

   \title{The GW-Universe Toolbox III: simulating joint observations of gravitational waves and gamma-ray bursts}

   %\subtitle{I. Overviewing the $\kappa$-mechanism}

   \author{Kai Hendriks
          \inst{1}\thanks{E-mail: kai.hendriks@gmail.com}
          \and
          Shu-Xu Yi\inst{2,1}
          \and
          Gijs Nelemans\inst{1,3,4}
          }

   \institute{Department of Astrophysics/IMAPP, Radboud University, P.O. Box 9010, 6500 GL Nijmegen, The Netherlands
         \and
             Key Laboratory of Particle Astrophysics, Institute of High Energy Physics, Chinese Academy of Sciences, 19B Yuquan Road, Beijing 100049, People’s Republic of China
            \and
            SRON, Netherlands Institute for Space Research, Niels Bohrweg 4, 2333 CA Leiden, The Netherlands
            \and
            Institute of Astronomy, KU Leuven, Celestijnenlaan 200D, B-3001 Leuven, Belgium
             }

   \date{\today}

% \abstract{}{}{}{}{} 
% 5 {} token are mandatory
 
  \abstract
  % context heading (optional)
  % {} leave it empty if necessary  
   {In the current multi-messenger astronomy era, it is important that information about joint gravitational wave (GW) and electromagnetic (EM) observations through short gamma-ray burst (sGRBs) remains easily accessible to each member of the GW-EM community. The possibility for non-experts to execute quick computations of joint GW-sGRB detections should be facilitated.}
  % aims heading (mandatory)
   {In this study, we construct a model for sGRBs and add this to the framework of the previously-built \texttt{Gravitational Wave Universe Toolbox} (\GWT or \Tb). We provide expected joint GW-sGRB detection rates for different combinations of GW detectors and high-energy (HE) instruments.}
  % methods heading (mandatory)
   {We employ and adapt a generic GRB model to create a computationally low-cost top-hat jet model suitable for the \GWT. With the \Tb, we simulate a population of binary neutron stars (BNSs) observed by a user-specified GW detector such as LIGO, Virgo, the Einstein Telescope (ET) or the Cosmic Explorer (CE). Based on the characteristics of each binary, our model predicts the properties of a resulting sGRB, as well as its detectability for HE detectors such as \textit{Fermi}/GBM, \textit{Swift}/BAT or GECAM.}
  % results heading (mandatory)
   {We report predicted joint detection rates for combinations of GW detectors (LIGO and ET) with HE instruments (\textit{Fermi}/GBM, \textit{Swift}/BAT, and GECAM). Our findings stress the significance of the impact that ET will have on the multi-messenger astronomy; where the LIGO sensitivity is currently the limiting factor regarding the number of joint detections, ET will observe BNSs at such a rate, that the vast majority of detected sGRBs will have a GW counterpart observed by ET. These conclusions hold for CE as well. Additionally, since LIGO can only detect BNSs up to redshift $\sim$0.1 where few sGRBs exist, a search for sub-threshold GW signals at higher redshifts using sGRB information from HE detectors has the potential to be very successful and significantly increase the number of joint detections. Equivalently, during the ET era, GW data can assist in finding sub-threshold sGRBs, potentially increasing e.g. the number of joint ET-\textit{Fermi}/GBM observations by $\sim$270\%. Lastly, we find that our top-hat jet model underestimates the number of joint detections that include an off-axis sGRB. We correct for this by introducing a second, wider and weaker jet component. We predict that the majority of joint detections during the LIGO/Virgo era will include an off-axis sGRB, making GRB170817A not as unlikely as one would think based on the simplest top-hat jet model. In the ET era, most joint detections will contain an on-axis sGRB.}
  % conclusions heading (optional), leave it empty if necessary 
   {}


   \maketitle
%
%-------------------------------------------------------------------

\section{Introduction}
\label{sec:intro}
Nearly five years after the first joint observation of a gravitational wave (GW) signal and a short gamma-ray burst (sGRB) from a binary neutron star (BNS) merger \citep{abbott_multi-messenger_2017,abbott_gw170817_2017,goldstein_ordinary_2017}, multi-messenger astronomy has established itself as a central part of modern astrophysics. The ability to observe astrophysical events both gravitationally and electromagnetically allows for unique insights into (astro)physical phenomena, potentially probing new fundamental understandings in physics. The implications of multi-messenger astrophysics stretch beyond GW physics, as this concept can be applied to many different areas within physics and astrophysics. For example, one can use joint detections to compute the Hubble constant (e.g. \cite{chen_distance_2021, chen_program_2021}), rule out or confirm certain binary progenitor models (e.g. \cite{eichler_nucleosynthesis_1989}), improve our understanding on neutron star (NS) interior and equation-of-state (EOS) (e.g. \cite{metzger_lessons_2019}), or carry out tests of general relativity \citep[e.g.][]{kim_prospects_2021, clark_prospects_2015}. The number of joint observations will only increase in the near future; BNS and black hole - neutron star (BHNS) mergers, the main candidates for joint detections due to their production of GW signals as well as electromagnetically sGRBs, will be observed more frequently. Current detectors such as aLIGO \citep{harry_advanced_2010} and Virgo \citep{acernese_advanced_2015} will eventually operate at their design sensitivities, and the planned Einstein Telescope (ET, \cite{punturo_einstein_2010}) and Cosmic Explorer (CE, \cite{reitze_cosmic_2019}) will observe GW sources at an unprecedented rate in this frequency band ($\sim$10 - 1000 Hz). Additionally, next to already established high-energy (HE) detectors such as \textit{Fermi}/GBM \citep{meegan_fermi_2009} and \textit{Swift}/BAT \citep{gehrels_swift_2004}, newer instruments like Insight-HXMT \citep{song_first_2022} and GECAM-GRD \citep{xiao_-ground_2022, zhang_energy_2019} (launched in 2017 and 2020 respectively) will prove extremely useful in the search of sGRBs as electromagnetic (EM) counterparts to GW observations.

%Given that the detection rate of binary neutron stars (BNSs) or black hole - neutron stars (BHNSs), i.e. GW sources that typically show EM signatures, by current GW detectors is relatively low, the extent to which questions such as these can be answered is currently limited. However, in the near future, this number will only increase.
% What do we wanna say in the intro?
% - MMA is booming and will only become bigger in future
% We should maybe explain what a sGRB is first, then talk about the birth of MMA, and how MMA will be growing and growing

%The mid-frequency (...) and low-frequency range (...), which may also be accompanied by EM counterparts, will be covered as well by the Laser Interferometer Space Antenna (LISA, ..) and pulsar timing arrays (PTAs, ...) respectively.
%Paragraph explaining the new things that joint observations can lead to:
%- H0
%- ruling out binary progenitor models (see Clark)
%- get maximum NS mass (Metzger)
%- understand jet breakout and probe opening angle (Clark)
%
%Paragraph about the importance of GRB
Previously, we built a open-source Python based software, the \texttt{Gravitational Wave Universe Toolbox} \citep[the \GWT or \Tb hereafter for short]{yi_gravitational_2021,yi_gravitational_2022} to make GW astrophysics easily accessible. The \GWT can be found at \url{gw-universe.org}, or downloaded as a Python package at \url{https://bitbucket.org/radboudradiolab/gwtoolbox}. The \GWT can be used to simulate observations with different kinds of GW detectors on various common GW source populations, among which are BNS and BHNS mergers. It can return a synthetic catalogue of BNS/BHNS, which could be detected with a user-specified GW detector, signal-to-noise (SNR) threshold and observation duration. For each event in the catalogue, the \Tb returns the masses of the binary, luminosity distance, orbital inclination and effective spin. We have now added\footnote{For now, our sGRB model is only available in the Python package of the \GWT. Eventually, it will be added to the website as well.} to this a sGRB model that allows the \Tb to return the properties of a potential sGRB from these sources as observed by a user-selected HE detector. This enables the user to study the detectability of these events as GRB with respect to a certain HE detector, and investigate joint GW-EM detection rates. This new functionality makes the \GWT a multi-messenger simulator, and paves the way towards more applications in studies such as prospects of joint-detections with different GW/HE detectors and sub-threshold strategies\footnote{By this we mean the targeted search for signals in GW data whose signal-to-noise ratio is below the typical detection threshold, using information from an observed GRB counterpart (or the other way around, i.e. using GW data to find sub-threshold observations in GRB datastreams).}, GRB physics and compact binary merger history. The first study is conducted in this work, while we will discuss in more detail several planned other projects in Section \ref{sec:summary}. Where this research is in a sense similar to a recent study by \cite{ronchini_perspectives_2022}, it should be noted that our project touches upon several slightly different aspects of the prospects of multi-messenger astronomy, meaning that these two studies can be viewed as complementary to one another.

%Especially within the current multi-messenger astronomy era, a tool such as this is extremely useful due to its wide range of possible applications: e.g. research into BNS/BHNS properties optimal for joint detections, redshift distributions of joint observations, sub-threshold detections, or future joint observation strategies.

In this paper, we describe the sGRB model that has been added to the \Tb, as well as investigate joint detection rates predicted by the upgraded \GWT. The paper is organised as follows: in Section \ref{sec:methods}, we will first give a brief introduction on how the \GWT gives the synthetic catalogue of BNS/BHNS. Then we describe how GRBs emission from a BNS/BHNS progenitor is simulated in our model, and how we determine the detectability of the GRB for a certain HE detector. In Section \ref{sec:results}, we first reproduce individual and population properties of historical GRBs, and by doing that we fix the parameters in the underlying population and GRB models. Subsequently, we use the simulation to predict future detection rates, with different combinations of GW/HE detectors, with sub-threshold joint-observation strategies, and specifically for off-axis sGRBs. In Section \ref{sec:summary} we summarise our findings and provide future prospects.
\section{Methods}
\label{sec:methods}
%\label{sec:methods}

\subsection{The \GWT: producing a synthetic GW catalogue}
Here we give a brief explanation on how a synthetic GW catalogue is generated in the \GWT. For a detailed description, we refer to \cite{yi_gravitational_2021}. For a certain GW detector, its noise power spectrum and the antenna pattern are denoted as $S_n(f)$ and $F_{+,\times}$ respectively. The $S_n(f)$ is either from literature if the GW detector is from a default list, or calculated with \texttt{FINESSE} \citep{brown_pykat_2020} if it has a user-customised configuration. The detector response to the GW from a compact binary merger is: 
\begin{equation}
    h(f)=F_+h_+(f)+F_\times h_\times(f),
\end{equation}
%\red{can this be varied or is it always IMRPhenomD?} (\blue{Shuxu: In principle it can be replaced by any approximant from pycbc. But it's not very convenient to do that. The FIM part is also specially designed for the parameters in IMRPhenomD.. So I'll say it's unchangeable for now.})
where $h_{+,\times}$ are the "plus" and "cross" polarisations of the GW waveform, which are calculated with a certain waveform approximant (e.g., \texttt{IMRPhenomD}, \cite{husa_frequency-domain_2016, khan_frequency-domain_2016}). The signal-to-noise ratio (S/N, SNR, or $\rho$) of the GW is calculated as:
\begin{equation}
        \rho^2=4\int^{f_{\rm{high}}}_{f_{\rm{low}}}\frac{|h^2(f)|}{S_{\rm{n}}(f)}df.
    \label{eqn:snrLVK}
\end{equation}

\begin{figure}
        \centering
        \includegraphics[width=.5\textwidth]{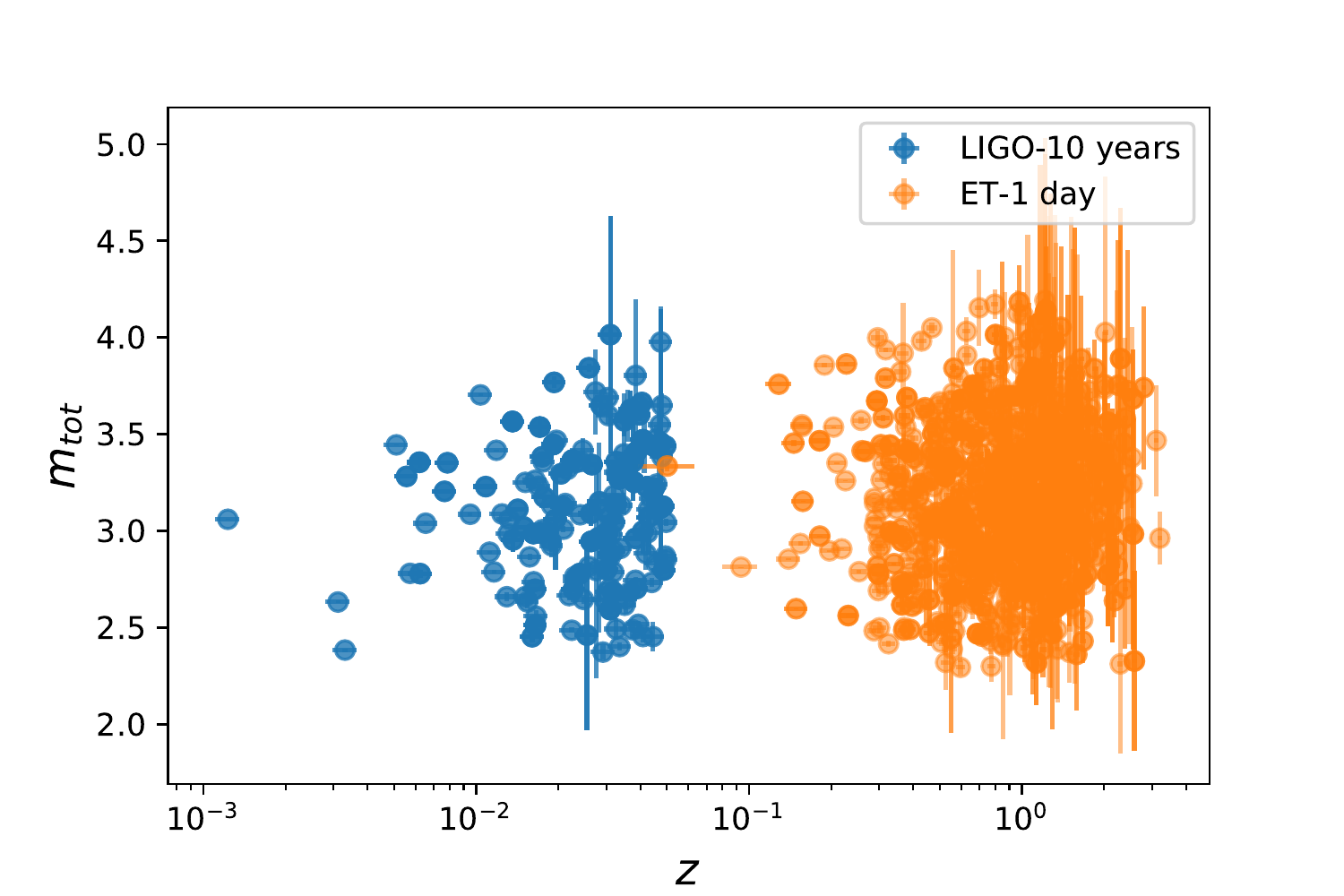}
        \caption{Redshift distribution of a simulated BNS merger catalogue from the \GWT for a LIGO observation time of 10 years (blue) and 1 day of ET observations (orange).}
        \label{fig:example1}
    \end{figure}

For a certain S/N threshold, we can define a probability of detection of a source as:
\begin{equation}
    \mathcal{D}(\mathbf{\Theta})=\oiint d\Omega d\Omega^\prime\mathcal{H}(\rho^2-\rho^2_\star)/(4\pi)^2,
\end{equation}
$\mathbf{\Theta}$ denotes the intrinsic parameters (e.g., masses and effective spin, depending on the waveform approximant) and the luminosity distance of the source, $\Omega$ and $\Omega^\prime$ denote the sky location angles ($\theta$, $\phi$), the inclination angle ($\iota$) and the polarisation angles ($\psi$); $\mathcal{H}$ is the Heaviside step function. With a merger rate model of the population, $\dot{n}(\mathbf{\Theta})$, the expected distribution of detectable sources is:
\begin{equation}
    N_{\rm{D}}(\mathbf{\Theta})=\frac{T}{1+z}\frac{dV_{\rm{c}}}{dz}\dot{n}(\mathbf{\Theta})\mathcal{D}(\mathbf{\Theta}),\label{eqn:detectables}
\end{equation}
and the expected number of detection is an integration of $N_{\rm{D}}(\mathbf{\Theta})$ over the possible parameter space. A synthetic catalogue is obtain by a MCMC sampling from $N_{\rm{D}}(\mathbf{\Theta})$. Fig. \ref{fig:example1} is an example of the simulated catalogues from the \GWT. The \GWT has one default parameterised BNS model (\texttt{pop-A}) and two BHNS population models (\texttt{pop-A} and \texttt{pop-B}). For a detailed description of these population models, we refer to \cite{yi_gravitational_2021}. The common hyper-parameters for BNS, BHNS-\texttt{pop-A} and \texttt{pop-B} are listed in Tab. \ref{tab:BNS_hyper} (including hyper-parameters for the GRB model, which will be introduced in Sec. \ref{sec:model}). For BNS, there are five extra hyper-parameters describing the mass distribution of the NS, and its effective spin. In our treatment, we assume no dependence of the GRB properties on these parameters. For BHNS-\texttt{pop-A}, there are 9 extra parameters describing the masses distribution of BH and NS, and their effective spin. The values of these hyper-parameters influence the fraction of BHNS system which passes the $R_{\rm{tid}}>R_{\rm{ISCO}}$ criterion (see Sec. \ref{sec:cond}). We list these hyper-parameters and their default values in Tab. \ref{tab:BHNS-popA}. The difference between BHNS-\texttt{pop-A} and \texttt{pop-B} is that the latter includes another Gaussian mass peak centered at $\sim 40\,M_\odot$ on the mass function of $\texttt{pop-A}$ to take into account the BH originated from pair-instability supernovae. 
\begin{table*}
    \centering
    \begin{tabular}{c|c|c}
    \hline
    \textbf{Hyper-parameter}      & \textbf{Description} & \textbf{Default value}\\
    \hline
    $R_{\rm{n}}$     & Normalisation factor of the cosmic merger rate of BNS merger & 50 Gpc$^{-3}$ yr$^{-1}$\\
    $\tau$ & the averaged time-delay between the binary star formation and BNS merger & 10 Gyr\\
    $\Delta\theta_{\rm{mean}}$ & Mean jet opening angle & 15$^{\circ}$\\
    $\Delta\theta_{\rm{std}}$ & Standard deviation of the jet opening angle & 4$^{\circ}$\\
    $\log r_{0,\rm{mean}}$ & Mean log GRB radius & 13 $\log\text{cm}$\\
    $\log r_{0,\rm{std}}$ &  Standard deviation of log GRB radius & 0.1 $\log\text{cm}$\\
    $\log E_{\rm{GRB,mean}}$ & Mean of total GRB energy & 49.3 $\log\text{ergs}$\\
    $\log E_{\rm{GRB,std}}$ & Standard deviation of total GRB energy & 0.5 $\log\text{ergs}$\\
    $\nu^\prime_{0,\rm{mean}}$&  the mean value of the reference frequency in the jet co-moving frame & 0.001 MeV\\
    $\nu^\prime_{0,\rm{std}}$& Standard deviation of the reference frequency in the jet co-moving frame & 0.00075 MeV\\
    $\gamma_{\rm{mean}}$ & Mean of the bulk Lorentz factor & 250\\
    $\gamma_{\rm{std}}$ & Standard deviation of the bulk Lorentz factor & 75\\
    $t_{d, std}$ & Standard deviation of the time delay $t_d$ & 0.4 s\\
    \hline
    \end{tabular}
    \caption{Common hyper-parameters for both BNS/BHNS originated GRB population model. Their default values are listed in the third column, with which, the historical observation can be optimally reconstructed.}
    \label{tab:BNS_hyper}
\end{table*}

\begin{table*}
\centering
\begin{tabular}{c|c|c}
\hline
\textbf{Hyper-parameter}      & \textbf{Description} & \textbf{Default value}\\
\hline
$m_{\rm{n,mean}}$ & The mean of NS masses & 1.4\,$M_\odot$ \\
$m_{\rm{n,std}}$ & The standard deviation of NS masses & 0.5\,$M_\odot$ \\
$m_{\rm{n,high}}$ & The higher limit of NS masses & 2.5\,$M_\odot$ \\
$\gamma$ & the power index of the mass function tail & 2.5 \\
$c$ & $c/\gamma+\mu$ is the peak mass of BH mass function & 15\\
$m_{\bullet,\rm{cut}}$ & the upper limit of BH mass & 95\,$M_\odot$ \\
$\chi_{\rm{eff}}$ & the standard deviation of the effective spin parameters & 0.1 \\
\hline
\end{tabular}
\caption{Extra hyper-parameters of the underlying BHNS population of model \texttt{pop-A}.}
\label{tab:BHNS-popA}
\end{table*}

\begin{figure}
    \centering
    \includegraphics[width=0.5\textwidth]{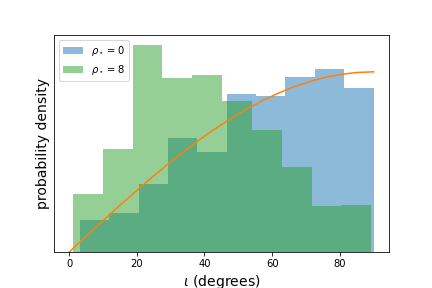}
    \caption{Inclination angle distribution of GW detection of different S/N threshold.}
    \label{fig:inc_dis}
\end{figure}

The catalogue simulated in this way does not include the orbital inclination angles, since the angles are marginalised when calculating the probability of detection. Since the inclination angle $\iota$ is a crucial property of GRBs, we want to recover $\iota$ from $\mathcal{D}(\mathbf{\Theta})$. We note that $\mathcal{D}(\mathbf{\Theta})$ is equivalent to the fraction in a uniform sample of $(\theta, \phi, \iota, \psi)$ that results in $\rho\ge\rho_{\star}$. We know:
\begin{equation}
    \rho\propto\sqrt{\left(\frac{1+\cos^2\iota}{2}\right)^2F_+^2+\cos\iota^2F_\times^2},
\end{equation}
where we define the right-hand side as $\bar{\rho}$. For a GW event in the catalogue, with a detection probability $\mathcal{D}$: we draw a sample of uniformly distributed $(\theta, \phi, \iota, \psi)$, and calculate the corresponding $\bar{\rho}$. We randomly select a $\Tilde{\rho}$ which is larger than the $(1-p)$-th quantile. The corresponding $\iota$ is assigned as the inclination angle of the source. In Fig. \ref{fig:inc_dis}, we plot the distribution of $\iota$ of the simulated detected sample of BNSs. We can find in the histogram that for $\rho_{\star}=0$, $\iota$ follows an isotropic distribution ($\sin\iota$). When $\rho_{\star}=8$, the distribution of $\iota$ peaks at around $\sim30^\circ$. This is a result of the increasing GW signal towards lower inclinations combined with the  $\sin\iota$ of the solid angle. 

\subsection{BNS and BHNS mergers that result in a sGRB}
\label{sec:cond}
%\blue{Shuxu: Here you can write something like in your thesis 2.1.1, but briefly. and then say: Here we assume that every BNS event will launch a propmt GRB. }
%To bridge the gap between gravitationally detectable properties and the occurrence of a sGRB, we discuss the conditions that the merger needs to satisfy in order to produce a sGRB.
In general, the merger remnant after a BNS or BHNS is only able to produce a sGRB if it is surrounded by a sizeable accretion disk powering a relativistic jet \citep[e.g.][]{bernuzzi_neutron_2020, fryer_fate_2015, zhang_physics_2018}. For BNS systems, in cases when the binary promptly collapses to a black hole (BH), it is known that the remnant satisfies this criterion. However, depending on the initial binary mass, different scenarios are possible for the final fate of the BNS merger remnant: apart from a BH, a BNS may form a short-lived hyper- or supermassive NS collapsing to a BH, or result in an infinitely stable supermassive NS \citep[e.g.][]{ciolfi_short_2018}. There still exists uncertainty regarding the ability of a NS remnant to contain an accretion disk and relativistic jet. However, recent observations as well as numerical evidence suggest that NS remnants may indeed launch jets resulting in a sGRB \citep{sarin_evolution_2020}. As such, in the \Tb, we assume that all of the simulated BNSs will launch a relativistic jet after merger, triggering a sGRB.

For a BHNS system, only in the case that the NS is swallowed by the BH before it is tidally disrupted, there will be no associated EM counterpart \citep{fryer_fate_2015}. Therefore, we impose a condition for the association of a prompt GRB with a BHNS as:
\begin{equation}
R_{\rm{tid}}>R_{\rm{ISCO}},
\label{eq:bhns_condition}
\end{equation}
where $R_{\rm{tid}}$ is the tidal radius, within which the NS will be tidally disrupted by the BH. It can be calculated with:
\begin{equation}
    R_{\text{tid}}=\big(\frac{M_\bullet}{M_{\rm{NS}}}\big)^{1/3}R_{\rm{NS}},
\end{equation}
 where $M_\bullet$ and $M_{\rm{NS}}$ are masses of the BH and NS respectively, and $R_{\rm{NS}}$ is the radius of the NS. $R_{\rm{ISCO}}$ is the radius of the Innermost Stable Circular Orbit (ISCO) around the BH, within which the NS is thought to be swallowed by the BH. The component masses of the BHNS are outputs from the GW catalogue of the \Tb. However, the $R_{\rm{NS}}$ depends on the mass-radius-relation of a NS. Therefore, in the \GWT, we build in four possible mass-radius-relations (MRRs): one relativistic Brueckner-Hartree-Fock EoS (MPA1) \citep{muther_nuclear_1987}, two relativistic mean-field theory EOSs (MS0, MS2) \citep{muller_relativistic_1996}, and one nonrelativistic potential model (PAL1) \citep{prakash_equation_1988}. In general, any tabulated MRR can be incorporated. %\red{[Gijs: this really needs a bit more explanation. Nobody will know what they mean. Also give the original reference for the equation of state, not a general paper!]}

\begin{figure}
    \centering
    \includegraphics[width=0.5\textwidth]{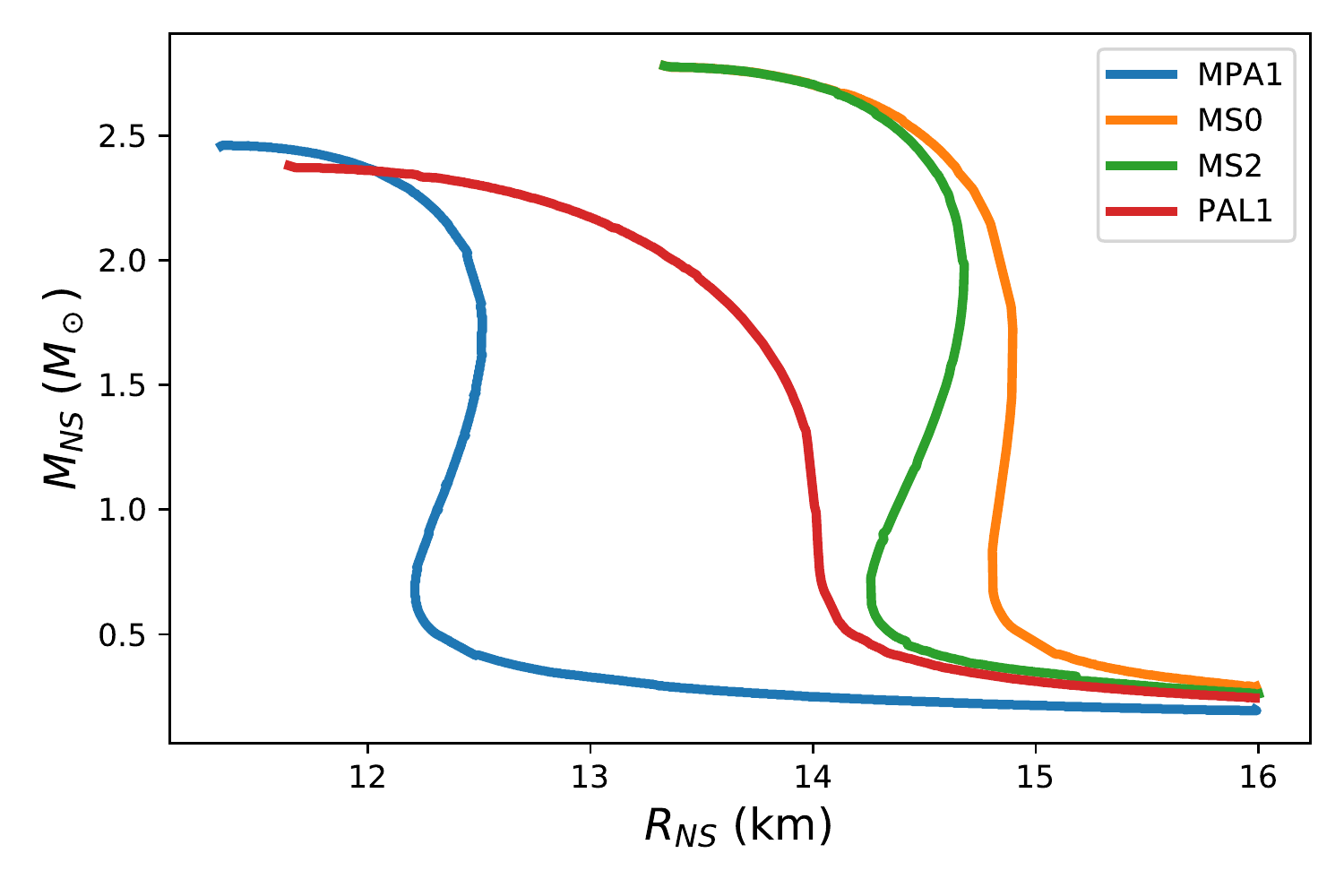}
    \caption{Built-in MRRs in the \GWT. We use the Brueckner-Hartree-Fock EoS (MPA1), two relativistic mean-field theory EOSs (MS0, MS2), and one nonrelativistic potential model (PAL1).}
    \label{fig:MRR}
\end{figure}

\subsection{The sGRB model}
\label{sec:model}
With a catalogue of GRB progenitors, we still need a GRB emission model so as to simulate the GRB from those progenitors. Here we employ and adapt a generalised geometrical model for GRBs from \cite{ioka_peak_2001}, which is independent of the details of the physics, and is therefore compatible with various physical GRB models. Its mathematical structure allows for potential future adaptations for an increased complexity and accuracy to be easy to implement. Additionally, it serves the purpose of the Toolbox: while it should recover the GRB with a reasonable accuracy, its simplicity keeps the computational cost low allowing for quick computations. The core quantity is the observed flux at the observed instance $T$:
\begin{equation}
    F_\nu(T)=\frac{\nu D}{\gamma\beta}\int^{2\pi}_0\phi\int^{\alpha_m}_0\alpha^2d\alpha\int^{\nu\gamma(1+\beta)}_{\nu\gamma(1-\beta)}\frac{d\nu^\prime}{\nu^\prime}\frac{j^\prime_{\nu^\prime}}{(1-\mu^2)^{3/2}},
    \label{eq:flux_generic}
\end{equation}
where $\gamma$ is the bulk Lorentz factor of the jet, $\beta=\sqrt{1-\gamma^2}$; $\alpha$, $\alpha_m$ and $\mu$ are geometry parameters describing the relative position between the emitting region and the observer. $j^\prime_{\nu^\prime}$ is the emissivity in the jet-comoving frame, which is assumed to be isotropic. For more details, see \cite{ioka_peak_2001} as well as \cite{woods_constraints_1999} \& \cite{piran_gamma-ray_1999}.

\begin{figure*}[t]
    \centering
    \includegraphics[width=\textwidth]{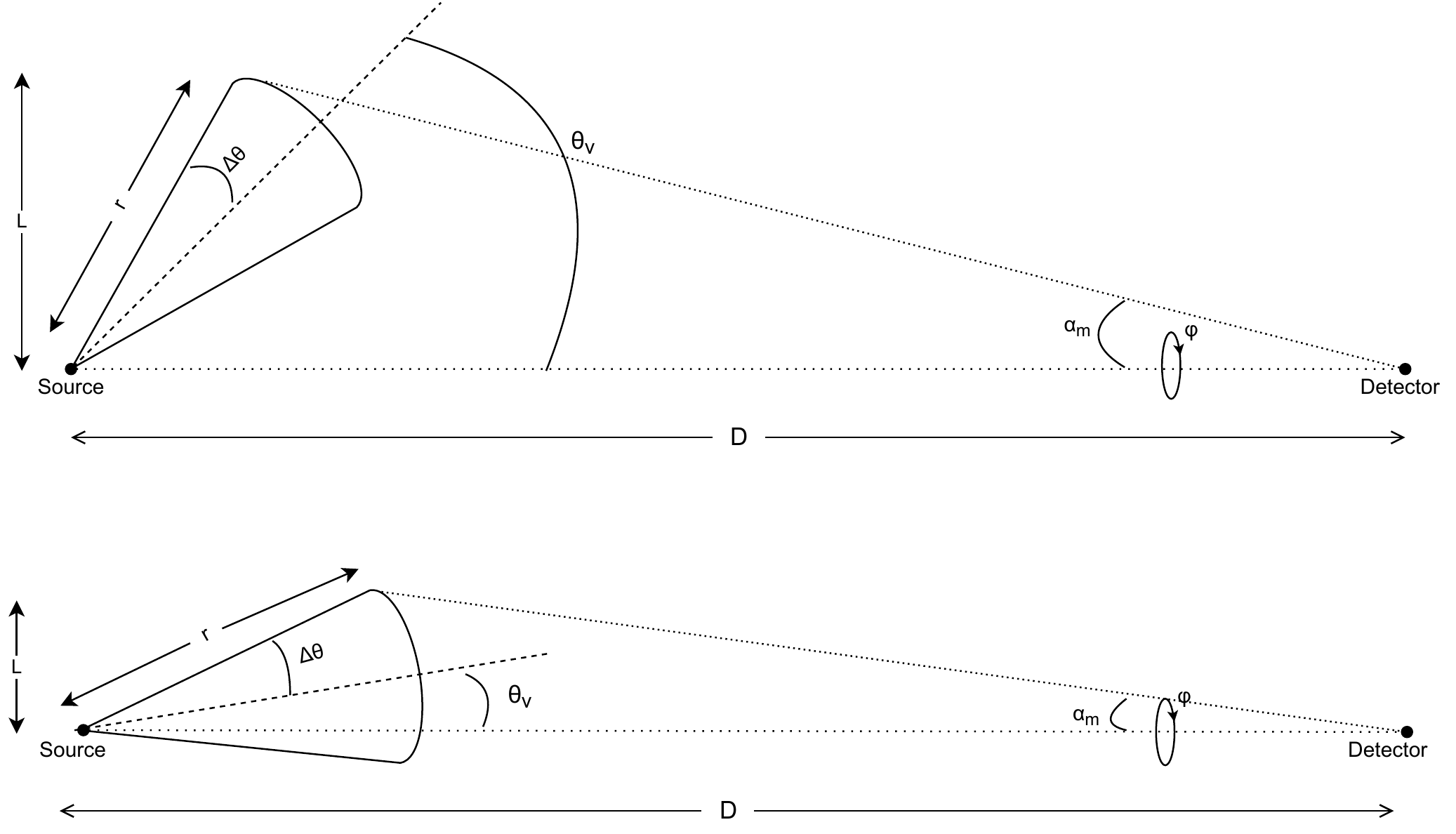}
    \caption{Schematic overview of the structure and geometry of our sGRB model, for a jet with opening angle $\Delta\theta$, viewing angle $\theta_v$ and distance to earth $D$. We depict an off-axis GRB in the top panel, while the bottom shows an on-axis one.}
    \label{fig:sgrb_structure}
\end{figure*}

The emissivity $j^\prime_{\nu\prime}$ is constructed as follows:
\begin{equation}
    j^\prime_{\nu^\prime} = A_0 f(\nu^\prime)\pi(t)\psi(r)\Pi(\Omega),
\end{equation}
where $\pi(t)$ and $\psi(r)$ are, respectively, the temporal and radial distribution of the emission, in the comoving frame. $\Pi(\Omega)$ is the angular structure of the jet, $A_0$ is a constant, and $f(\nu^\prime)$ denotes the spectrum in the comoving frame. Here, we employ the most simplified case, where a top-hat jet with instantaneous emission at time $t_0$ and radius $r_0$ is assumed. We get:
\begin{align}
    &\pi(t) = \delta(t - t_0),
    \label{eq:pi_t}
    \\
    &\psi(r) = \delta(r - r_0),
    \label{eq:psi_r}
    \\
    \begin{split}
        &\Pi(\Omega) = H(\Delta\theta - |\theta - \theta_v|)\\
        &\qquad H\left[\cos\phi - \left(\frac{\cos\Delta\theta - \cos\theta_v \cos\theta}{\sin\theta_v \sin\theta}\right)\right].
        \label{eq:pi_omega}
    \end{split}
\end{align}
$\Delta\theta$ and $\theta_v$ are the half-opening angle and viewing angle of the jet, respectively. For the spectrum $f(\nu^\prime)$, we use the shape of the Band function \citep{band_batse_1993}:
\begin{equation}
    f(\nu^\prime)=\big(\frac{\nu^\prime}{\nu^\prime_0}\big)^{1+\alpha_B}\big(1+\frac{\nu^\prime}{\nu^\prime_0}\big)^{(\alpha_B-\beta_B)/s}.
\end{equation}
We set $\alpha_B=-1$, $\beta_B=-2.2$, and $s=1$, as these are typical values resulting in a spectrum whose shape matches the Band spectrum \citep{preece_batse_2000,ioka_can_2018}. With the assumptions of Eqs. \ref{eq:pi_t}-\ref{eq:pi_omega}, Eq. \ref{eq:flux_generic} becomes
\begin{equation}
    F_\nu(T)=\frac{2cA_0r_0\gamma^2}{D^2}\frac{\Delta\phi(T)f\big(\nu\gamma(1-\beta\cos\theta(T))\big)}{\bigg(\gamma^2\big(1-\beta\cos\theta(T)\big)\bigg)^2}
    \label{eqn:FvT}.
\end{equation}
We visualise the geometry of our setup in Fig. \ref{fig:sgrb_structure} for an off-axis (top) and on-axis GRB (bottom).

In equation (\ref{eqn:FvT}), the constant $A_0$ is related to the total energy of the GRB ($E_{\rm{GRB}})$. In order to find the relation between $A_0$ and $E_{\rm{GRB}}$, we conduct the following derivation. In the comoving frame, the radiating energy of the ejecta is:
\begin{equation}
    E^\prime_{\rm{GRB}}=\int j^\prime dV\int d\nu^\prime \int d\Omega \int dt,\label{eqn:Egrb}
\end{equation}
where $j^\prime$ is the emissivity in the comoving frame, $\int dV$, $\int d\nu^\prime$, $\int d\Omega$ and $\int dt$ are integrals over volume, frequencies, emission directions and time respectively. Assuming isotropy of emission in the comoving frame, the direction integral gives $4\pi$, and with the assumption of instantaneous emission at an infinitely thin layer with a top-hat jet, the volume and time integrals give
$$\int dV \int dt = 2\pi(1-\cos\Delta\theta)r^2_0.$$ The frequency integral of the Band spectrum gives $$\int f(\nu^\prime) d\nu^\prime= -A_0\nu^\prime_0\frac{s^2\Gamma(\frac{\alpha}{2}+1+\frac{2}{s})\Gamma(-\frac{\beta}{s}+1-\frac{2}{s})}{(\alpha+2)(\beta+2)|s|\Gamma(-\frac{\beta-\alpha}{2})}.$$
%% _____________________________________________________________
Therefore, equation (\ref{eqn:Egrb}) becomes:
\begin{equation}
    E^\prime_{\rm{GRB}}=8\pi^2(1-\cos\Delta\theta)r^2_0A_0\mathcal{I}\nu^\prime_0,
\end{equation}
where 
$$\mathcal{I}=-\frac{s^2\Gamma(\frac{\alpha}{2}+1+\frac{2}{s})\Gamma(-\frac{\beta}{s}+1-\frac{2}{s})}{(\alpha+2)(\beta+2)|s|\Gamma(-\frac{\beta-\alpha}{2})},$$
which is 6.64 when $s=1$, $\alpha=-1$ and $\beta=-2.2$. Note that energy is not Lorentz invariant, and in the observer frame, the energy transformed to:
\begin{equation}
    E_{\rm{GRB}}=\gamma E^\prime_{\rm{GRB}}.
\end{equation}
So, we have 
\begin{equation}
    E_{\rm{GRB}}=A_0\gamma8\pi^2(1-\cos\Delta\theta)r^2_0A_0\mathcal{I}\nu^\prime_0.
\end{equation}
Equivalently:
\begin{equation}
    A_0=\frac{E_{\rm{GRB}}}{\gamma8\pi^2(1-\cos\Delta\theta)r^2_0A_0\mathcal{I}\nu^\prime_0}.
\end{equation}
It should be noted that $E_{\text{GRB}}$ depends on the mass of the accretion disk surrounding the merger remnant as well as the accretion efficiency, both of which depend on the binary parameters \citep{bernuzzi_neutron_2020,fryer_fate_2015}. Where efforts have been made to connect these quantities (e.g. \cite{radice_binary_2018,bernuzzi_accretion-induced_2020,nedora_mapping_2020}), as of yet the exact relationship between the binary properties and GRB energy is not known. As such, taking $E_\text{GRB}$ as a direct free parameter instead of indirectly through e.g. the binary mass is currently the safest option. More so, the total GRB energy is typically well-approximated by assuming a standard reservoir clustered around some central energy value \citep{frail_beaming_2001, zhang_physics_2018}.
%Please note that I didn't include a factor 2 to account for symmetric jet of both sides. That's because in observation, people report the ``beaming-corrected Energy" without taking that factor into account either. I think that's because you're not sure about whether the jet is indeed symmetric on both side. 

%In equation (\ref{eqn:FvT}), $r_0$ is the distance from the central engine, where the prompt GRB is emitted (or sometimes called the GRB radius). In the internal shock model with uniform, isotropic emission and a conical jet, the GRB radius is equal to the internal shock radius: $R_{GRB}\approx2\Gamma^2c\Delta t$. $\Delta t$ is typically of the order $10^{-2}$. Typical values for the internal shock radius are therefore around $10^{12} - 10^{13}$ cm \citep{zhang_physics_2018}.

Note that equation (\ref{eqn:FvT}) assumes an instantaneous emission from one location $r=r_0$ and $t=r_0/c$. The corresponding lightcurve is a single pulse which rises sharply at $T=0$ for on-axis cases. In reality the lightcurve is a superposition of multiple pulses, resulting in a more gradual rise. We recover this feature by implementing a temporally extended jet launch: every GRB pulse is launched with a different delay time $t_{\rm{d}}$ (in the source frame) relative to the merger, and $t_{\rm{d}}$ follows an extended probability distribution. The lightcurve can be obtained by convolving the original lightcurve with the probability distribution of (the red-shifted) $t_{\rm{d}}$, for which we use a Gaussian with a width $t_{\rm{d},std}(1+z)$, centred at $3t_{\rm{d},std}(1+z)$.

%\teal{Kai: I think we should also mention here that the model is simply one pulse, from a fixed time and radius, and that we extended it a bit by assuming that there are a number of pulses around t0, all with some delay time, and that all of them added together form the lighcurve. We can also add the delay time to the list of intrinsic parameters.}
%The first one is related to the high-latitude curvature effect. Photons from higher latitude with respect to the line-of-sight arrive later, which creates the decaying lightcuve. Additionally, the photons that arrive later (from higher latitude and angle from the line-of-sight) have a slightly softer spectrum. Models that include this effect require an inner shock radius of $\geq 10^{15}$ cm to fit with observations. Secondly, if we assume that all optical and gamma-ray emission comes from the same radius ($R_{GRB}$), we can use the synchrotron self-absorption (SSA) theory to derive a lower limit on $R_{GRB}: 10^{14}$cm.
As a summary, our GRB requires the following parameters.\\
Intrinsic parameters:
\begin{itemize}
    \item $\gamma$: the bulk Lorentz factor of the jet;
    \item $r_0$: the radius from the central engine, where the gamma-ray emission happens;
    \item $E_{GRB}$: the total energy released into prompt gamma-ray, in the observer frame, this is related with the normalization parameter $A_0$;
    \item $\alpha_B$, $\beta_B$, $s$ and $\nu^\prime_0$: intrinsic spectrum parameters;
    \item $\Delta\theta$: the half opening angle of the jet cone;
    \item $t_{\rm{d,std}}$ the standard deviation of the intrinsic jet launch delay time distribution.
\end{itemize}
Extrinsic parameters:
\begin{itemize}
    \item $\iota$: the viewing angle of the observer, or equivalently, the inclination angle of the binary orbital plane; 
    \item $D$ or $z$: the luminosity distance from the source to the observer, or equivalently, the red-shift of the GRB.
\end{itemize}

For each GRB in the \Tb, the $\Delta\theta$, $\log r_0$, $\log E_{\rm{grb}}$, $\nu^\prime_0$, $\gamma$ are assigned as Gaussian randoms. The corresponding means and standard deviations are $\Delta\theta_{\rm{mean}}$ and $\Delta\theta_{\rm{std}}$, $\log r_{0,\rm{mean}}$ and $\log r_{0,\rm{std}}$, $\log E_{\rm{grb},\rm{mean}}$ and $\log E_{\rm{grb},\rm{std}}$, $\nu^\prime_{0,\rm{mean}}$ and $\nu^\prime_{0,\rm{std}}$, $\gamma_{\rm{mean}}$ and $\gamma_{\rm{std}}$ respectively. We summarise these hyper-parameters for the underlying GRB population in Tab.\ref{tab:BNS_hyper}. The inclination angle $\iota$ and the distance $D$ are taken directly from the GW part of the \GWT.
%As an example for the GRB model, we calculate here the time-evolving spectrum of a source with parameters like GRB170817a, with equation (\ref{eqn:FvT}). We use $E_{\gamma}=8\times10^{49}$\,ergs, $\Delta\theta=25^\circ$, $\theta_v=29^\circ$, $r_0=10^{13}$\,cm, $D=40$\,Mpc, $\nu^\prime_0=10$\,keV. The simulated time-evolving spectrum is plotted in Fig. \ref{fig:170817spec_lag}. We then integrate the time-evolving spectrum in the energy from 10 keV to 25 MeV, which corresponds to the \textit{Fermi}/GBM energy range, to obtain the lightcurve. The simulated lightcurve is plotted in Fig. \ref{fig:170817_lc}. From the lightcurve we can obtain its burst duration $T_{90}=1.1$ s, the peak energy $E_{\rm{peak}}=100$ keV and the fluence $S=2\times10^{-7}$ ergs/cm$^2$, which are all compatible with the real observation. If we put the same source 100 times further at 4\,Gpc, but observed on-axis, then the simulation gives: $T_{90}=0.8$\,s, $E_{peak}=5800$\,keV, $S=9.6\times10^{-9}$\,ergs/cm$^2$/s.

%\begin{figure}
%    \centering
%    \includegraphics[width=0.5\textwidth]{images/170817_spec_lag.pdf}
%    \caption{The simulated time-evolving spectrum of GRB170817a.}
%    \label{fig:170817spec_lag}
%\end{figure}
%\begin{figure}
%    \centering
%    \includegraphics[width=0.5\textwidth]{images/170817_light_curve.pdf}
%    \caption{The simulated lightcurve of GRB170817 as observed with \textit{Fermi}/GBM.}
%    \label{fig:170817_lc}
%\end{figure}

\begin{figure}
\centering
\begin{subfigure}{0.49\textwidth}
    \centering
    \includegraphics[width=\textwidth]{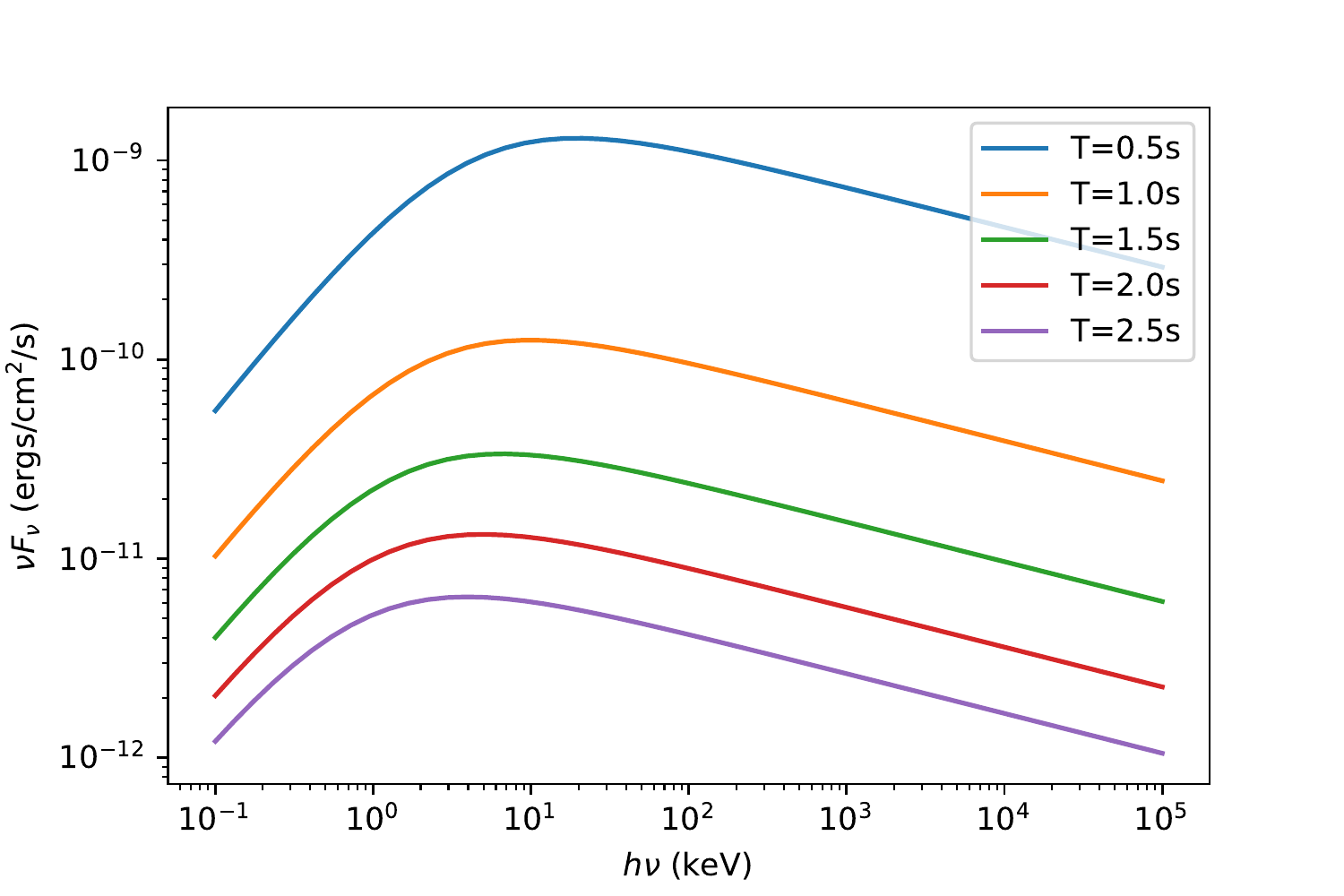}
    \caption{The simulated time-evolving spectrum of the on-axis sGRB GRB131004A.}
    \label{fig:spectrum_onaxis}
\end{subfigure} \\
\begin{subfigure}{0.49\textwidth}
    \centering
    \includegraphics[width=\textwidth]{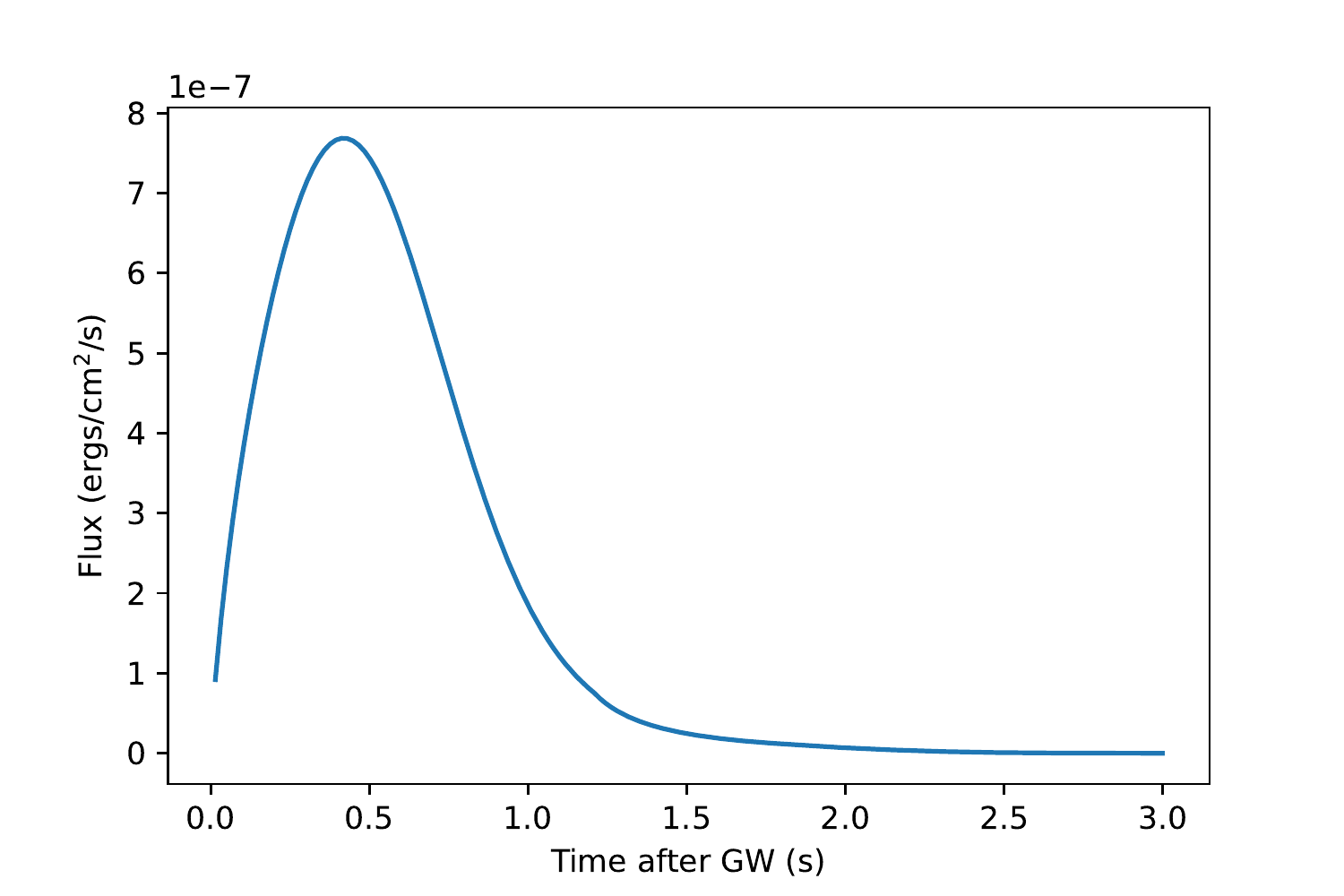}
    \caption{The simulated lightcurve of the on-axis GRB131004A, within an energy range of 10-1000 keV.}
    \label{fig:lightcurve_onaxis}
\end{subfigure}
\caption{Characteristics of a GRB131004A-like on-axis GRB, according to the \GWT. We use the following hyper-parameters: $\gamma = 100$, $\log(r_0/\text{cm)}=12.93$, $\log(E_{\text{GRB}}/\text{ergs})=49.8$, $\nu_0^\prime=0.7$keV, $\Delta\theta=15^\circ$, $\theta_v=10^\circ$, and $z=0.717$.}
\label{fig:onaxis}
\end{figure}

\begin{figure}
\begin{subfigure}{0.49\textwidth}
    \centering
    \includegraphics[width=\textwidth]{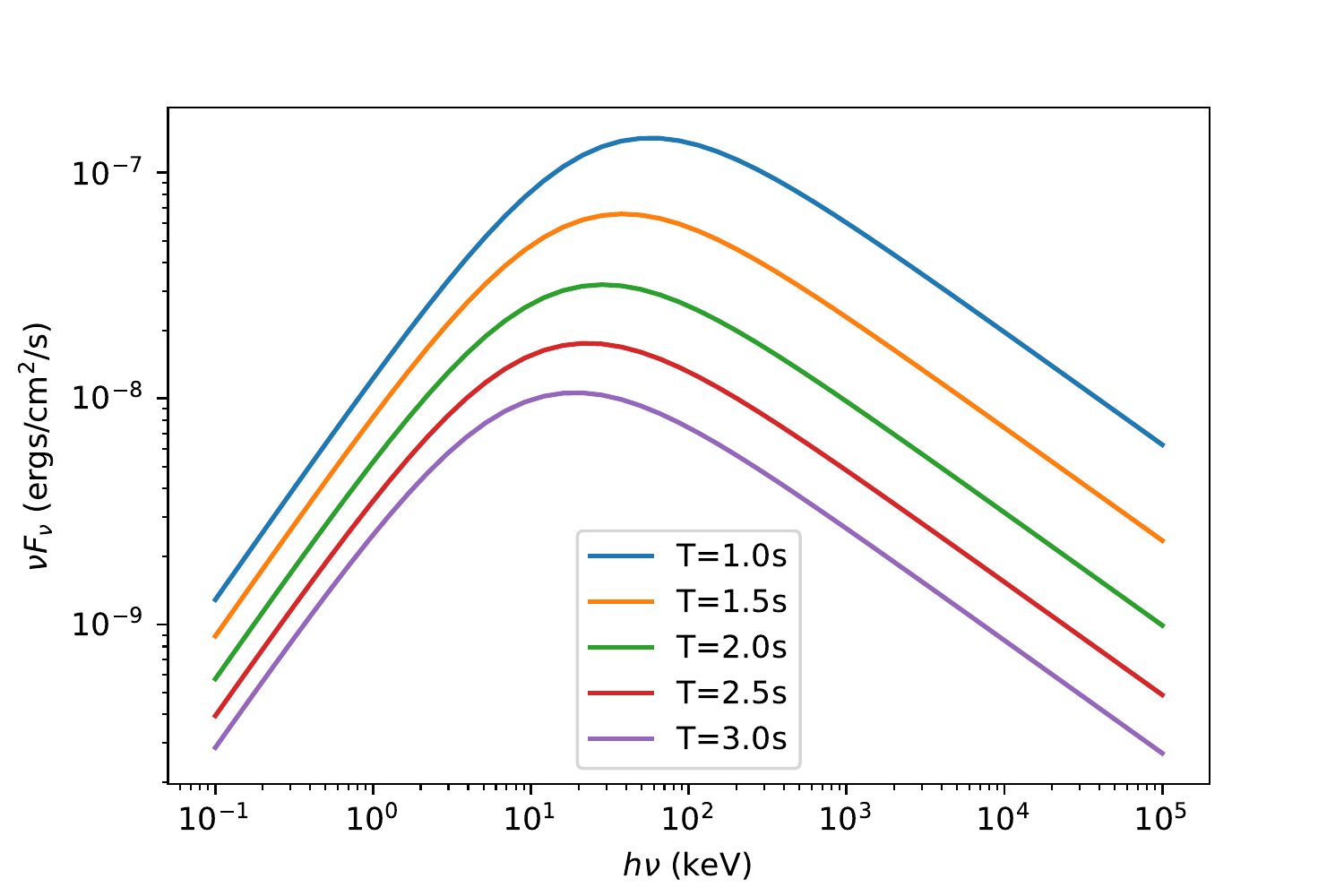}
    \caption{The simulated time-evolving spectrum of the off-axis sGRB GRB170817A.}
    \label{fig:spectrum_offaxis}
\end{subfigure}
\\
\begin{subfigure}{0.49\textwidth}
    \centering
    \includegraphics[width=\textwidth]{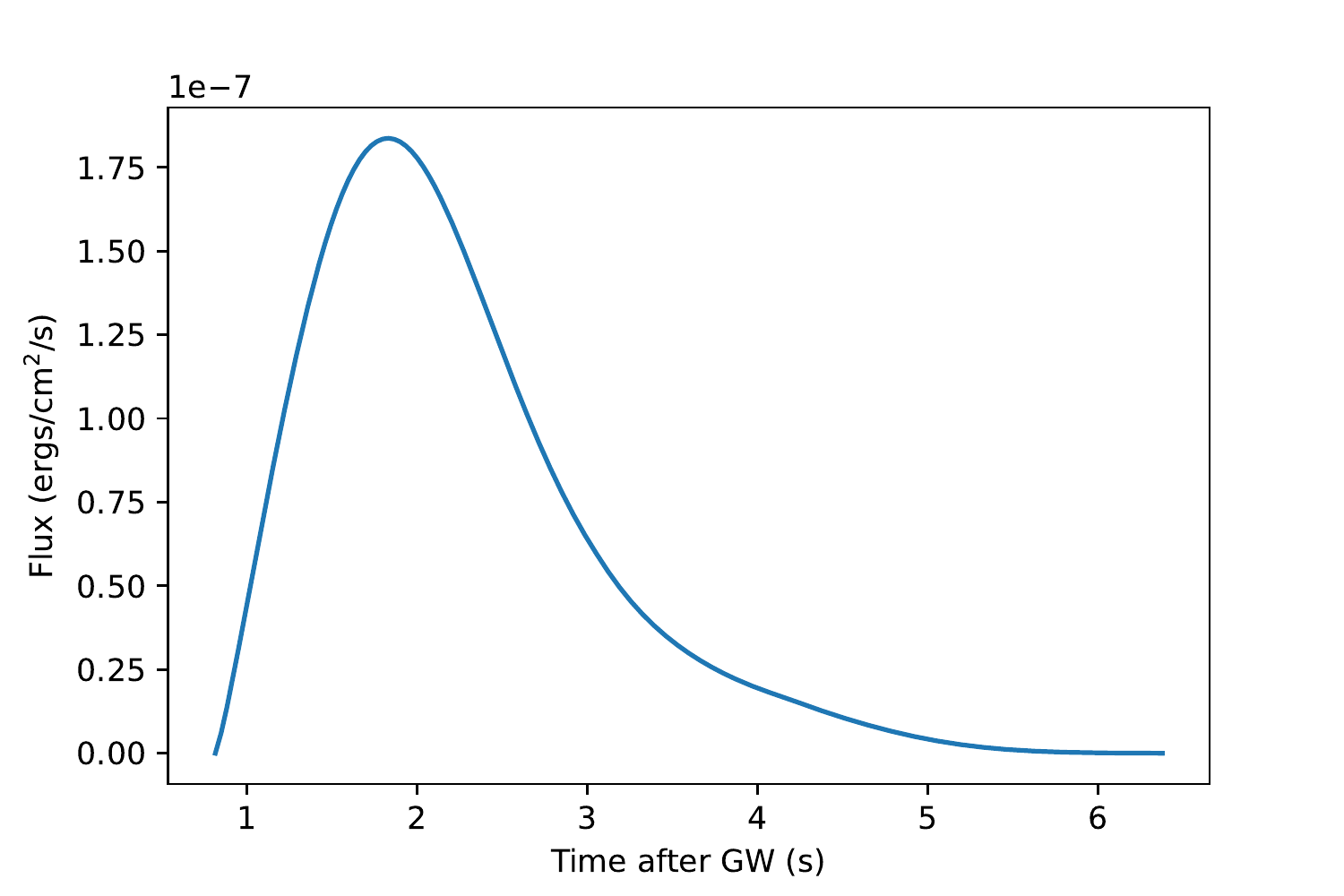}
    \caption{The simulated lightcurve of the off-axis GRB170817A, within an energy range of 10-1000 keV.}
    \label{fig:lightcurve_offaxis}
\end{subfigure}
\caption{Characteristics of a GRB170817A-like off-axis GRB, according to the \GWT. We use $\gamma = 100$, $\log(r_0/\text{cm}) = 13.03$, $\log(E_{\text{GRB}}/\text{ergs})=49.65$, $\nu_0^\prime=8$ keV, $\Delta\theta=15^\circ$, $\theta_v=18.75^\circ$, and $z=0.0099$.}
\label{fig:offaxis}
\end{figure}

%\g{I would consider moving this to section 2.3. In addition, there are no observations in the plots, which makes it difficult to compare. Some things (T90, peak fluxes and Epeak) you could plot in the figures, isn't it? I would also put 7+8 and 8+9  in a single figure (two plots above each other)} \teal{Kai: I've changed the plots into two figures. In theory we could add flux observations at different points in time to the lightcurves, but I am not sure how much that would add. To me the figures are just there to show generally what the spectrum and lightcurve looks like, and from them we take the observables T90, 64-ms peak flux and Epeak and compare those with observations. Or is that not enough? What I do think we should still add is how exactly we extract these observables from the lightcurve/spectrum. I am fine with moving these graphs+explanations to 2.3, that would maybe make the flow in this section a bit better.}
To illustrate and visualise the model, in Figs. \ref{fig:onaxis} \& \ref{fig:offaxis} we show the time-evolving spectrum and lightcurve of an on- and off-axis sGRB respectively. The time-evolving spectrum $F_\nu(t)$ can be calculated with the method described above. The lightcurve, either in units of phs/s/cm$^2$ or ergs/s/cm$^2$, can be calculated with:

\begin{equation}
    L_{\rm{ph}}(t)=\int\frac{F_\nu(t)}{h\nu}d\nu,
    \label{eq:lightcurve_ph}
\end{equation}
or 
\begin{equation}
    L_{\rm{ergs}}(t)=\int F_\nu(t)d\nu, 
    \label{eq:lightcurve_ergs}
\end{equation}
respectively, with the integral over the photon frequencies corresponding to the energy range of the detector. In Fig. \ref{fig:onaxis}, we attempt to recover the properties of the on-axis sGRB GRB131004A, observed by \textit{Fermi}/GBM \citep{von_kienlin_fourth_2020,gruber_fermi_2014,von_kienlin_2nd_2014,bhat_third_2016}. With the hyper-parameters listed in Fig. \ref{fig:onaxis}, our model finds: $T_{90}=0.96$s (compared to an observed 1.152s), a fluence (10-1000 keV) of $5.64 \times 10^{-7}$ (observed $5.099 \times 10^{-7}$) ergs/cm$^2$, a 1-sec and 64-ms peak flux of respectively 5.85 (6.8) and 8.25 (9.8) ph/cm$^2$/s, and $E_{\text{peak}}=118$ (observed 118 as well) keV. We also show the evolving spectrum and lightcurve according to our model in the same energy range, for the off-axis GRB170817A observed by \textit{Fermi}/GBM \citep{goldstein_ordinary_2017} in Figs. \ref{fig:spectrum_offaxis} \& \ref{fig:lightcurve_offaxis}. Here, we find $T_{90}=2.55$s
(compared to an observed $2.0 \pm0.5$s), a fluence of $3.16 \times 10^{-7}$ (observed $2.8\pm0.2 \times 10^{-7}$) ergs/cm$^2$, a 64-ms peak flux of 3.00 (compared to $3.7\pm0.9$) ph/cm$^2$/s, and an $E_{\text{peak}}$ of 130 keV (observed $185\pm62$) keV. Overall, these predictions match the observations quite well for both GRBs. Still, for GRB170817A, we needed to choose quite an atypical value for $\nu_0^\prime$ in order to obtain somewhat realistic observables. This indicates that our top-hat jet model may contain several limitations which become exposed for off-axis GRBs, specifically regarding the angular dependence of the GRB spectrum; for a more detailed discussion, see Sec. \ref{sec:summary} and \cite{ioka_spectral_2019}. Nonetheless, keeping in mind that we expect the majority of GRB observations to be on-axis, if its main limitation is an inaccurate computation of off-axis spectral peaks, our model is accurate enough to serve the purpose of the \GWT. The aim of the \Tb is not to perfectly match all observations made by HE detectors; it should simply be able to recover the majority of the observations reasonably well, with low computational effort. This allows the user to make quick computations of joint detectabilities.

\subsection{GRB detectability against HE detectors}
\begin{table*}[t]
        \centering
        \begin{threeparttable}
        \begin{tabular}{l l l l l}
            \hline
            \textbf{Detector}& \textbf{Energy Range (keV)} & \textbf{Flux Limit} & \textbf{FoV (percentage)} & $\mathbf{f_c}$\\
            \hline
            Compton-BATSE & 50-300 & 0.3\text{phs/cm$^2$/s}& 25\% & 1 \\
            \textit{Swift}/BAT         & 15-150 & 1.5 \text{phs/cm$^2$/s}& 10\% & 2       \\
            \textit{Fermi}/GBM         & 10-1000 & 2.28 \text{phs/cm$^2$/s}& 60\% & 2\\
            %Konus  & 20-15000  & $10^{-7}$ ergs/s/cm$^2$& 80\% & 1\\
            \textit{Insight}-HE\tnote{1} & 30-250 & $2\times10^{-8}$ ergs/s/cm$^2$&  60\% & 1\\
            GECAM\tnote{2} & 15-5000 & $1.8\times10^{-8}$ ergs/s/cm$^2$& 30\% & 1\\
            %SVOM/GRM & 30-5000 & $2\times10^{-8}$ ergs/s/cm$^2$& 20\% & 1\\
            %SVOM/ECLAIRs & 4-250 & $2\times10^{-8}$ ergs/s/cm$^2$& 20\% & 1\\
            \hline
        \end{tabular}
        \begin{tablenotes}
        \item[1] While we include \textit{Insight}-HE in our list here, we leave this instrument out of our analysis; as its published GRB catalogue is currently incomplete, we are not able to extract an effective flux threshold at which it operates at the present time. Similarly to \textit{Fermi}/GBM and \textit{Swift}/BAT, using the theoretical threshold does not lead to accurate detection rates for sGRBs.
        \item[2] The designed FoV of GECAM is 100\%, however, due to a power problem, the actual on-orbit FoV had been only about $\sim30\%$ for the past year. During the preparation of this manuscript, we learnt that the FoV has been restored to $\sim60\%$ recently from a private conversation. 
        \end{tablenotes}
        \end{threeparttable}
        \caption{Default parameters for built-in HE detectors. Please note: the \textit{Swift}/BAT and \textit{Fermi}/GBM flux limits as well as the \textit{Fermi}/GBM energy range differ from their theoretical specified values. We follow \cite{coward_swift_2012} and realise that it is more difficult for a sGRB to be detected as it needs to produce a significant signal over a short amount of time. As such, we use the lowest value for the peak flux in the sGRB catalogues (\cite{lien_third_2016} for \textit{Swift}/BAT, \cite{von_kienlin_fourth_2020,gruber_fermi_2014,von_kienlin_2nd_2014,bhat_third_2016} for \textit{Fermi}/GBM) of each of these instruments, in the shortest time accumulation window (20 ms for \textit{Swift}/BAT, and 64ms for \textit{Fermi}/GBM where flux for the latter was given within 10-1000 keV). More HE detectors will be added to the \GWT in the future.}\label{tab:HEdetectors}
    \end{table*}
    
\begin{figure*}
\centering
\tikzstyle{process} = [rectangle, draw, text centered, minimum height=2em]
\tikzstyle{connector} = [draw, -latex]
\tikzstyle{decision} = [diamond, draw, text centered, minimum height=2em]
\tikzstyle{terminator} = [rectangle, draw, text centered, rounded corners, minimum height=2em]
    \begin{tikzpicture}[node distance = 3cm]
        \node[process,fill=blue!20](event){Event from catalogue};
        \node[process, right of =event,fill=blue!20](FvT){$F_{\nu}(T)$};
        \node[process, right of =FvT,fill=blue!20](LC){Lightcurve};
        \path [connector] (event) -- (FvT);
        \path [connector] (FvT) -- (LC);
        \node[decision, fill=blue!20, right of =LC](sens){$F_{\rm{max}}+F_{\rm{bkg}}\ge F_{\rm{limit}}$?};
        \node[terminator, fill=red!20, above of=sens](false){detected=False};
         \node[decision, fill=blue!20, right of =sens](FoV){In FoV?};
        \node[terminator, fill=green!20, below of=FoV](true){detected=True};
        \path [connector] (LC) -- (sens);
        \path [connector] (sens) -- (FoV);
        \path [connector] (FoV) -- (true);
        \path [connector] (sens) -- (false);
        \path [connector] (FoV) |- (false);
        \node[draw=none] at (8.5, 2) (no) {No};
        \node[draw=none] at (11.5, 2) (no2) {No};
        \node[draw=none] at (10.7, 0.5) (yes) {Yes};
        \node[draw=none] at (4.3, 0.3) (int){$\int^{\nu_{low}}_{\nu_{up}} d\nu$};
        \node[draw=none] at (11.5, -1.5) (yes2) {Yes};
    \end{tikzpicture}
\caption{The flowchart showing the sGRB detection test pipeline of BNSs.}
\label{fig:flow-ns}
\end{figure*}
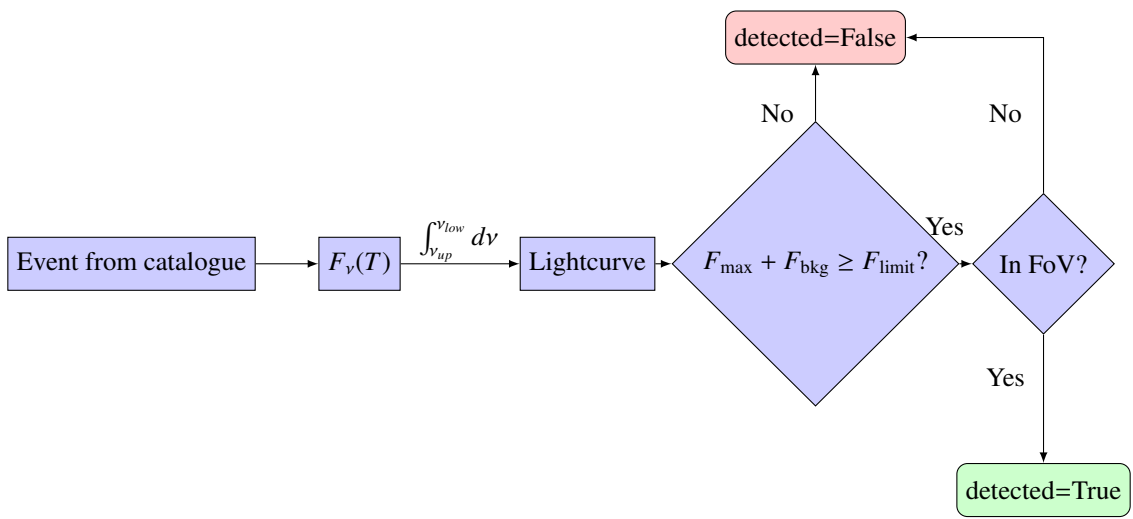

To decide whether a simulated sGRB is detectable by a selected HE instrument, the \Tb computes the peak of the lightcurve (using Eq. \ref{eq:lightcurve_ph} or \ref{eq:lightcurve_ergs}), and adds to this peak some typical background flux within the specified detector energy range. A GRB is detected if the sum of the GRB flux and background flux is above the detection threshold of the HE instrument. For \textit{Swift}/BAT sGRBs, a typical background is 0.3 ph/s/cm$^2$ \citep{barthelmy_burst_2005}, which, with a flux threshold of 1.5 ph/s/cm$^2$ \citep{coward_swift_2012}, leads to a detection if the total flux is $>5$ times that of the background. For simplicity, we assume the same factor of 5 for other detectors, and since we know the effective threshold of these other HE instruments, we can set the background accordingly. Even if a GRB is bright enough to be observed by a certain detector, it may still remain undetected, for either it is not in the detector's field-of-view (FoV), or it does not pass the detector's complex trigger algorithm. We treat this by introducing two parameters for each detector: i) the FoV as a percentage of the full sky, and ii) a factor $f_c$ to account for the correction due to the trigger algorithm. For each GRB that passes the sensitivity threshold, we compare a random variable $u$ in [0,1] with $\text{FoV}/f_c$. If $u<\text{FoV}/f_c$, we consider the GRB detected, otherwise we flag it undetected. In Figs. \ref{fig:flow-ns} \& \ref{fig:flow-nsbh}, we summarise the detection algorithm for a BNS and BHNS respectively. In Tab. \ref{tab:HEdetectors}, we list the parameters for our built-in HE detectors.
%\red{[Gijs: I find this a bit confusing. Isn't the background the thing that determines the detection, so I think it is clearer if you say that a total flux $>5$ times the background is assumed to be detectable. This seems quite high I must say. Or did I misunderstand something?} \teal{[Kai: I have now rewritten this part a bit to explain it more along the lines you suggest. Where 5 may seem quite high, Coward (2012) justifies this as follows: "Because SGRBs occur over a short duration, it is more difficult (compared to long bursts) to produce a significant signal above background". So in practice, the effective threshold for sGRBs is quite a bit higher than the theoretical threshold. For Swift/BAT, this becomes 1.5 ph/s/cm2.]}

\tikzstyle{process} = [rectangle, draw, text centered, minimum height=2em]
\tikzstyle{connector} = [draw, -latex]
\tikzstyle{decision} = [diamond, draw, text centered, minimum height=2em]
\tikzstyle{terminator} = [rectangle, draw, text centered, rounded corners, minimum height=2em]
 \begin{figure*}
 \centering
    \begin{tikzpicture}[node distance = 3cm]
        \node[process,fill=blue!20](event){Event from catalogue};
        \node[decision, fill=blue!20, right of=event]at (1,0)(tidal){Tidally disrupted?};
        \node[terminator, fill=red!20, above of=tidal](false){detected=False};
        \node[decision, fill=blue!20, right of=tidal]at (6,0)(BNSpipe){\small{Detected with BNS pipeline?}};
        \node[terminator, fill=green!20] at (9,-3.5)(true){detected=True};
        \path[connector](event)--(tidal);
        \path[connector](tidal)--(BNSpipe);
        \path[connector](BNSpipe)|-(false);
        \path[connector](BNSpipe)--(true);
        \path[connector](tidal)--(false);
        \node[draw=none] at (3.5, 2) (no) {No};
        \node[draw=none] at (7, 2.5) (no2) {No};
        \node[draw=none] at (9.8, -2.8) (yes) {Yes};
        \node[draw=none] at (6.1, 0.5) (yes2) {Yes};
    \end{tikzpicture}
\caption{The flowchart showing the sGRB detection test pipeline of BHNSs.}
\label{fig:flow-nsbh}
\end{figure*}
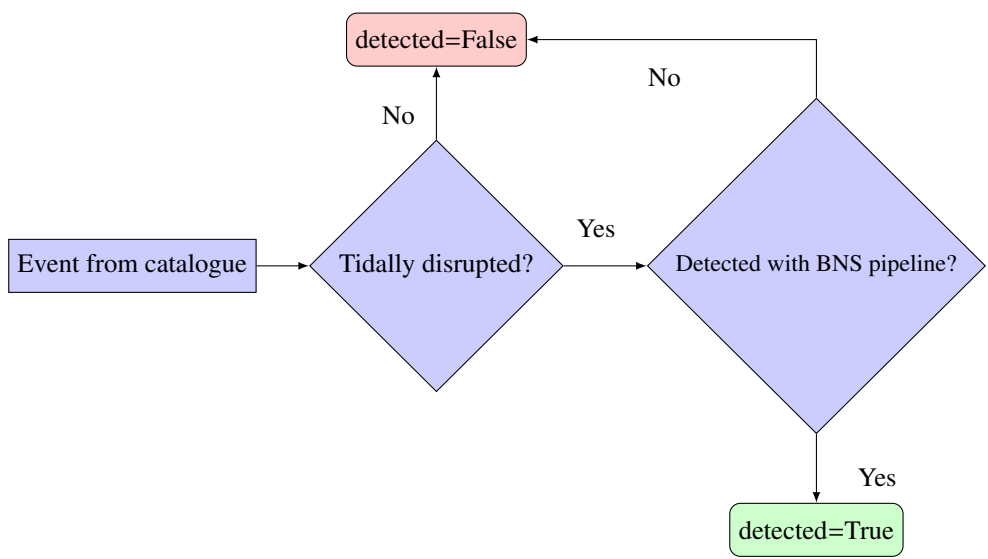

%\subsection{GRB population model} \red{[Gijs: I found this section a bit hard to read. Maybe it is an idea to discuss the population parameters in section 2.1 and the ones for the ones for the sGRB model in 2.3 and remove this section?]} \teal{Kai: that could work, I have moved the GRB model parameters part to 2.3, it's probably best if Shuxu takes care of the population parameters as he knows how to write it down most accurately.}
%For each GRB progenitor in the catalogue, the extrinsic parameters are sampled from the underlying BNS/BHNS population model. The \GWT has one default parameterised BNS model (\texttt{pop-A}) and two BHNS population models (\texttt{pop-A} and \texttt{pop-B}). For a detailed description of these population models, we refer to \cite{yi_gravitational_2021}. For the intrinsic parameters, we assign them empirically in the following way. For each GRB, the $\Delta\theta$, $\log r_0$, $\log E_{\rm{grb}}$, $\nu^\prime_0$, $\gamma$ are assigned as Gaussian randoms. The corresponding means and standard deviations are $\Delta\theta_{\rm{mean}}$ and $\Delta\theta_{\rm{std}}$, $\log r_{0,\rm{mean}}$ and $\log r_{0,\rm{std}}$, $\log E_{\rm{grb},\rm{mean}}$ and $\log E_{\rm{grb},\rm{std}}$, $\nu^\prime_{0,\rm{mean}}$ and $\nu^\prime_{0,\rm{std}}$, $\gamma_{\rm{mean}}$ and $\gamma_{\rm{std}}$ respectively. We summarise these hyper-parameters for the underlying GRB population in Tab.\ref{tab:BNS_hyper}. 
\section{Results and discussion}
\label{sec:results}
\subsection{Comparing the GRB model with observations}
%\blue{Shuxu: I remember that the BHNS contribution is negligible (most of them have $R_{\rm{t}}<R_{\rm{ISCO}}$). Can you give some numbers about it can conclude that we just focus on BNS pregenitors? }
Before investigating the \GWT GRB observations, we comment on our population parameters. In the \GWT  we use $R_0 = 50$ Gpc$^{-3}$ yr$^{-1}$ and $\tau = 10$ Gyr for the BNS population, and $R_0 = 10$ Gpc$^{-3}$ yr$^{-1}$ and $\tau = 10$ Gyr for the BHNS population. Both are in accordance with the merger rate density estimates from GW observations (\cite{abbott_population_2021} and \cite{abbott_observation_2021} respectively).
In this study, we assume that all generated sGRBs originate from BNSs. This assumption is justified as follows: the \Tb finds that the GRB-launching BHNS population (i.e. satisfying Eq. \ref{eq:bhns_condition}) in the universe ($\sim$2300 in 1 yr) is about 17 times smaller than its BNS equivalent ($\sim$40000 per year), and is therefore negligible for our purposes. Our model predicts $\sim$11 GRBs Gpc$^{-3}$ yr$^{-1}$ whose jet is directly in our line-of-sight (i.e. on-axis). This agrees with a previously predicted rate of 8$^{+5}_{-3}$ Gpc$^{-3}$ yr$^{-1}$ \citep{coward_swift_2012}. For a LIGO-O3 configuration, the \GWT finds a GRB detection rate of $\sim$0.68 per year, which we deem reasonably accurate as compared to the 1 BNS that was detected during O3 \citep{abbott_gw190425_2020}.

\begin{figure*}[t]
\centering
\begin{subfigure}[h]{0.49\linewidth}
        \includegraphics[width=\textwidth]{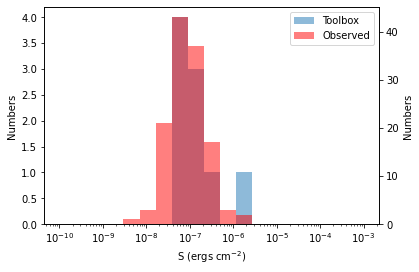}
        \caption{}
        \label{fig:swift_s}
    \end{subfigure}
~
\begin{subfigure}[h]{0.49\linewidth}
        \includegraphics[width=\textwidth]{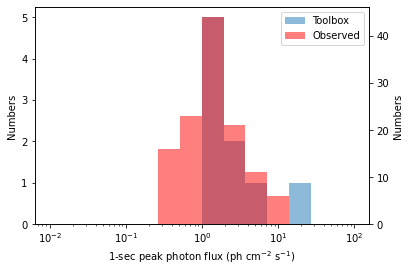}
        \caption{}
        \label{fig:swift_fmax}
\end{subfigure}\\
\begin{subfigure}[h]{0.49\linewidth}
        \includegraphics[width=\textwidth]{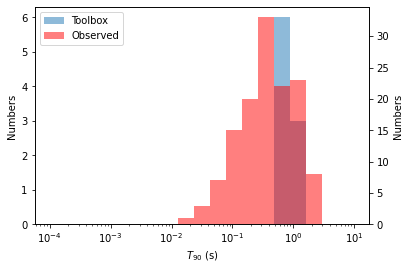}
        \caption{}
        \label{fig:swift_t90}
    \end{subfigure}
~
\begin{subfigure}[h]{0.49\textwidth}
        \includegraphics[width=\textwidth]{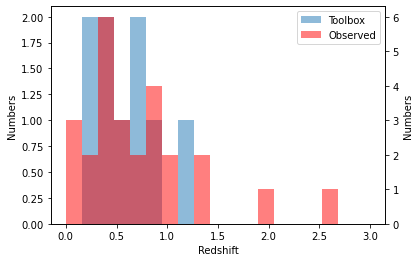}
        \caption{}
        \label{fig:swift_z}
\end{subfigure}\\
\begin{subfigure}[h]{0.49\textwidth}
        \includegraphics[width=\textwidth]{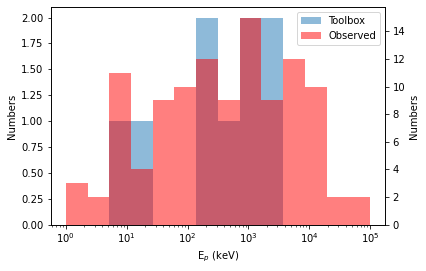}
        \caption{}
        \label{fig:swift_Ep}
\end{subfigure}
\caption{Fluence, 1-second peak photon flux, T90 and redshift histograms. We compare the \GWT simulation of \textit{Swift}/BAT observing sGRBs for 1 year (blue) with all sGRBs in the \textit{Swift}/BAT GRB catalogue (red) \citep{lien_third_2016}.}
\label{fig:swift_hist}
\end{figure*}

To verify how realistically the \GWT can recover the observed GRB population, we let it observe BNSs for 1 year with ET, using a SNR threshold of 0. This is an approximation to all BNS mergers in the universe within 1 year. Due to computational cost, we choose to not use a longer observing time. While this could lead to a potentially low number simulated sGRBs to compare with observations, we regard this as unproblematic as our goal is to only recover observations reasonably well. We constructed a detected GRB population from this by observing the BNSs with \textit{Swift}/BAT. In Fig. \ref{fig:swift_hist}, we show in blue several observables of this population. We compare these with the full observed \textit{Swift}/BAT GRB catalogue for GRBs with $T_{90}\leq$2s \citep{lien_third_2016}. 

First of all, we find that apart from the intrinsic GRB rate matching with \cite{coward_swift_2012}, we are also able to successfully reproduce actual observation rates for all detectors. In the case of \textit{Swift}/BAT depicted in Fig. \ref{fig:swift_hist}, we find $\sim$8 sGRBs per year, which is in line with the observed rate of 8 sGRB yr$^{-1}$ \citep{lien_third_2016}. It should be noted that we manually chose our correction factor $f_c$ such, that the resulting detection rate would (roughly) coincide with observations. Therefore, this comparison should be seen as nothing more than a check that this calibration was successful. Concerning other observables, it is especially important that we are able to recover the observed redshift distribution to at least some degree of accuracy, as the predictions for joint detection rates that we will make further along in this paper heavily rely on an accurate sGRB redshift distribution. Looking at Fig. \ref{fig:swift_hist}, we observe that in fact most histograms (redshift, fluence, and $E_p$) show a good match. Only the histograms for the duration $T_{90}$ and 1-second peak flux seem to indicate that a part of the observed population cannot be reproduced by the \GWT. In the case of $T_{90}$, sGRBs shorter than $\sim$0.3 seconds are not generated by the \Tb. The cause of this is our manually implemented time delay between merger and jet launch, quantified by the parameter $t_{d,std}$. Since we set this standard deviation as well as the corresponding mean to a fixed number ($0.4\times(1+z)$s and $1.2\times(1+z)$s respectively), there cannot be significant variations in $T_{90}$. Still, since the the duration is not nearly as decisive for the detectability of the GRB as e.g. the fluence or the redshift, this is not problematic for our purposes. Somewhat similarly, the 1-sec peak photon flux shows no detections below $\sim$1 ph cm$^{-2}$ s$^{-1}$. This is a direct result of our choice to approximate the complex \textit{Swift}/BAT triggering algorithm by setting a relatively high threshold (a 20-ms peak photon flux of 1.5 phs cm$^{-2}$ s$^{-1}$) and introducing a correction factor $f_c$. Where a more accurate representation would result in the inclusion of more low-flux (and low-fluence) sGRBs, this would be too complex for the (current version of) the \GWT. Nonetheless, the accuracy of our recovered redshift distribution clearly does not visibly suffer from the lack of low-energy sGRBs so for the purposes of this paper, we also deem this limitation acceptable.
%\g{the next part is not clear. Do you mean that the difference in the sample size is not important? You can say that more clearly. I would then also discuss the different plots, saying if match is good and if not why and why we don't think this is a problem. Where does the 8/yr come from? Better to discuss that first, saying that apart from the intrinsic match of the rate with Coward etc, you also roughly recover the actual detection rate.}

%Where it should be noted that it is hard to perform a quantitative analysis of the performance of the \Tb in this regard due to the low number of GRBs \textit{Swift}/BAT observes in a year, such an analysis is in fact not necessary for the \GWT for the same reasons as in the previous paragraph. As such, from Fig. \ref{fig:swift_hist} we conclude that the extent to which the GRB population parameters, including the detection rate of $\sim$8 sGRBs yr$^{-1}$, are recovered is acceptable for the purposes of the \Tb.

\subsection{Joint detection rates}
\begin{figure}
    \centering
    \includegraphics[width=0.5\textwidth]{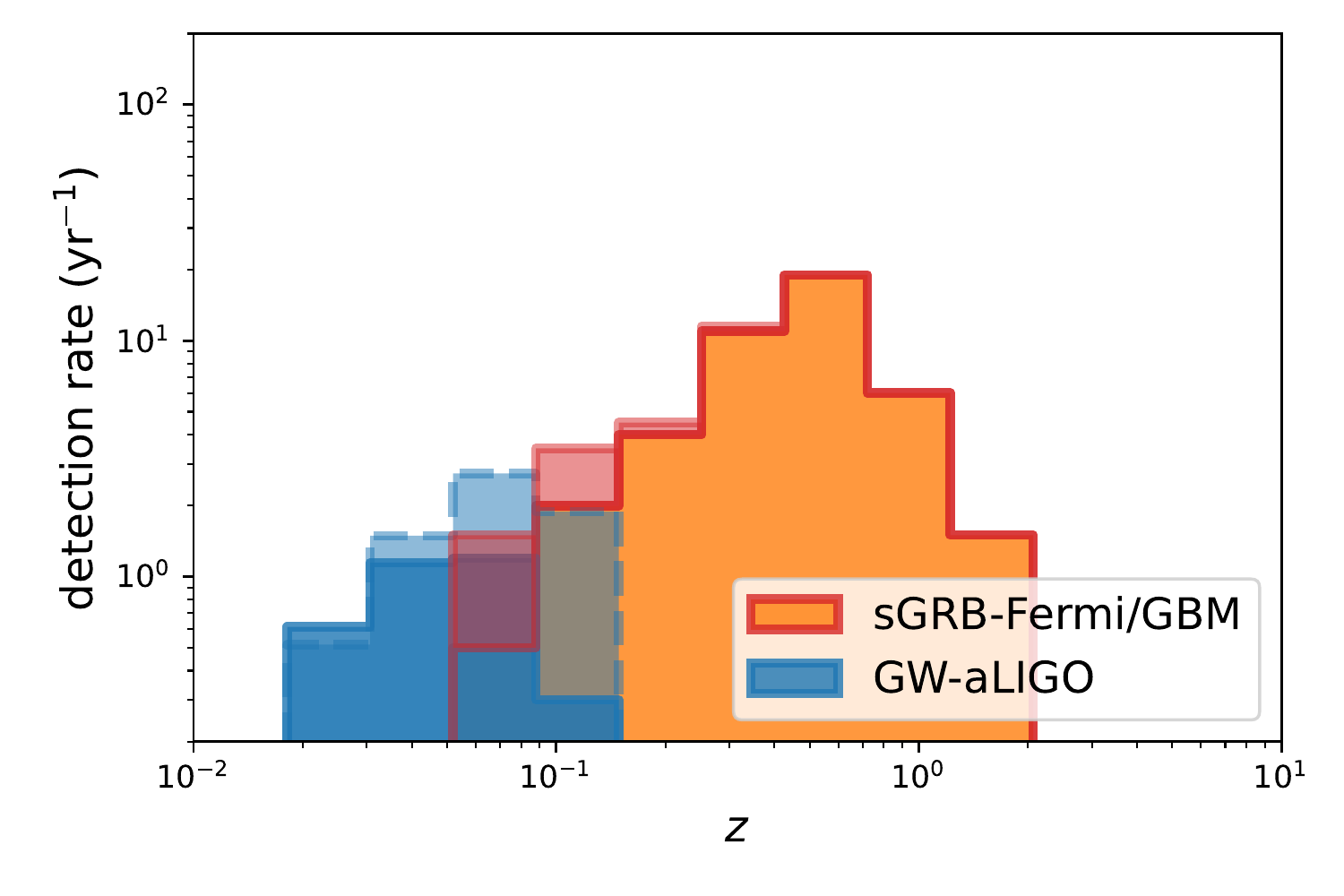}
    \caption{Redshift distribution of BNSs through detection of GWs and sGRBs. In blue, we show the distribution for LIGO (design sensitivity), where the shaded region with the dotted line represents the distribution for those GW sources with SNR between 6 and 8. In orange, we plot the distribution of \textit{Fermi}/GBM GRBs (the same distribution as in Figs. \ref{fig:ET_HE}-\ref{fig:halfET_HE}), where the red region corresponds to the off-axis GRBs, generated with the addition to our model as described in Sec. \ref{sec:offaxis}.}
    \label{fig:LIGO_HE}
\end{figure}

\begin{figure}
    \centering
    \includegraphics[width=0.5\textwidth]{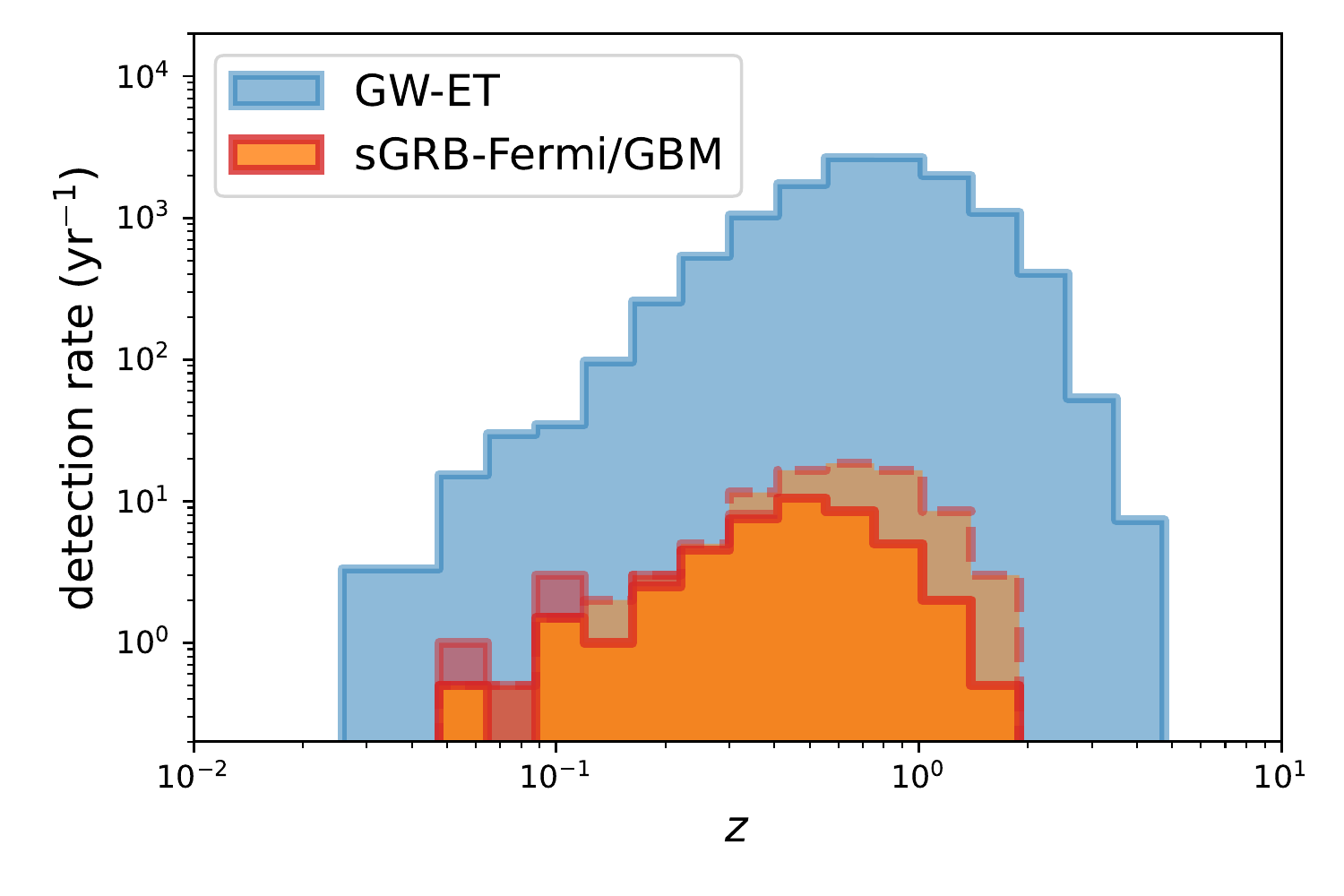}
    \caption{Redshift distribution of BNSs through detection of GWs and sGRBs. In blue, we show the distribution for ET. Again, in orange, we plot the distribution of \textit{Fermi}/GBM GRBs (same distribution as Figs. \ref{fig:LIGO_HE}, \ref{fig:LIGOO3_HE} \& \ref{fig:halfET_HE}), where the red region corresponds to the off-axis GRBs, generated with the addition to our model as described in Sec. \ref{sec:offaxis}. The orange region with dotted line represents those GRBs with a peak flux between 50\% and 100\% of the \textit{Fermi}/GBM detection threshold.}
    \label{fig:ET_HE}
\end{figure}

\begin{figure}
    \centering
    \includegraphics[width=0.5\textwidth]{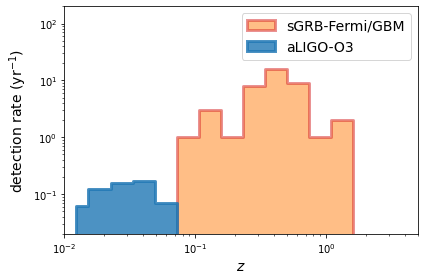}
    \caption{Redshift distribution of LIGO-O3 BNSs (blue), and \textit{Fermi}/GBM BNSs measured through sGRBs (orange). Once again, the \textit{Fermi}/GBM distribution is the same as in Figs. \ref{fig:LIGO_HE}, \ref{fig:ET_HE} \& \ref{fig:halfET_HE}.}
    \label{fig:LIGOO3_HE}
\end{figure}

\begin{figure}
    \centering
    \includegraphics[width=0.5\textwidth]{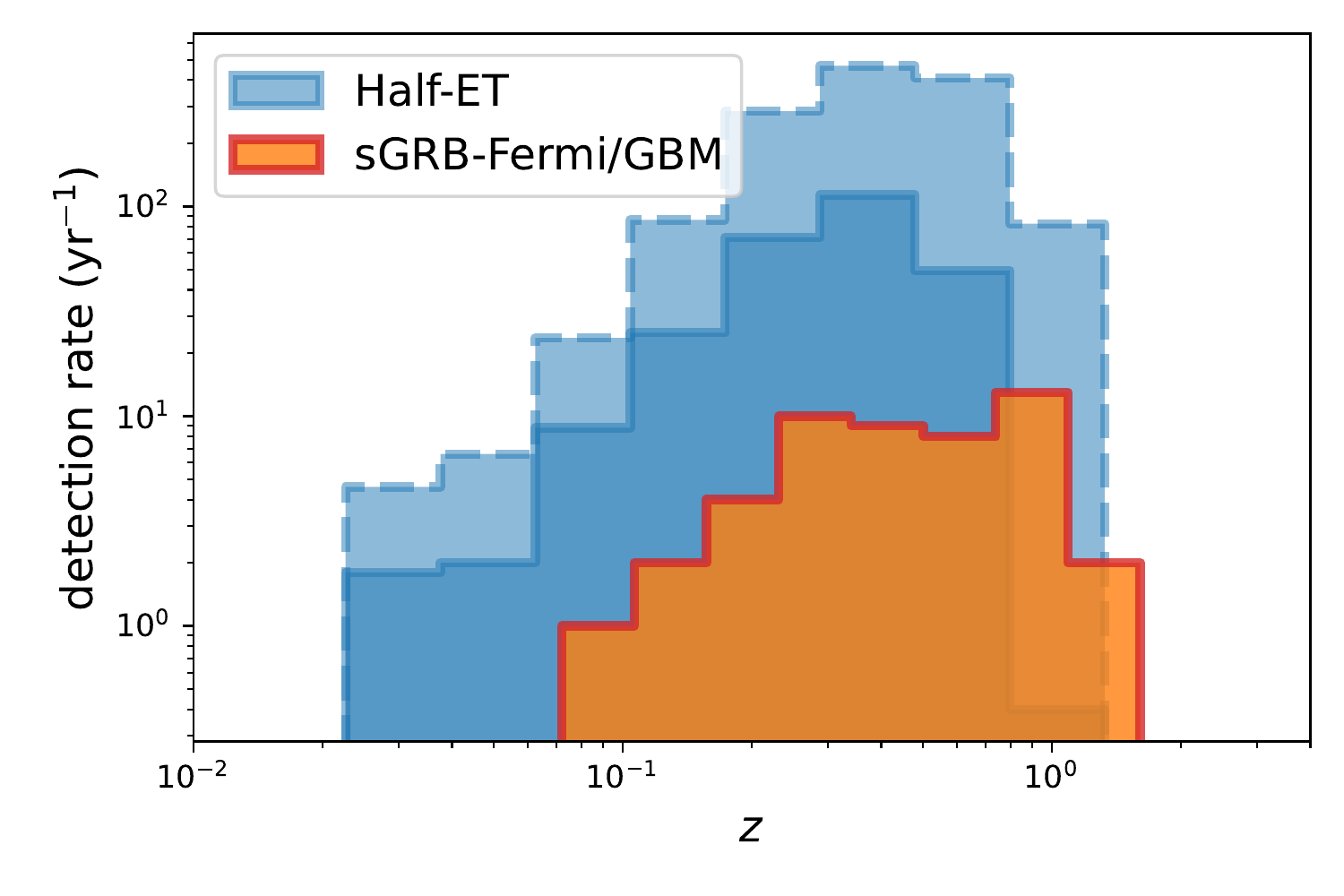}
    \caption{Redshift distribution of BNSs detected with a "half-ET" configuration (blue) which has a sensitivity halfway between LIGO and ET, and \textit{Fermi}/GBM BNSs measured through sGRBs (orange, same distribution as in Figs. \ref{fig:LIGO_HE}-\ref{fig:LIGOO3_HE}). Also here, the shaded region with the dotted line corresponds to those sources with a SNR between 6 and 8. SNR8: 2038, SNR6: 6019 in 5 yrs.}
    \label{fig:halfET_HE}
\end{figure}

In Tab. \ref{tab:GW_sub}, we show the number of GW detections per year for LIGO (design sensitivity) and ET, as well as joint GW-sGRB detections for four different HE detectors. We give numbers for 3 different SNR thresholds for the GW detection. In this Section, we will only discuss observations with the standard SNR of 8, whereas sub-threshold detections with SNR 7 or 6 will be addressed in Sec. \ref{sec:sub}. While expected joint rates for ET are in line with other predictions (e.g. \cite{ronchini_perspectives_2022}), our predictions for LIGO are lower than those in other studies (e.g. \cite{howell_joint_2019, clark_prospects_2015}). This is due to the low number of off-axis GRB observations according to our model. We further discuss this in Sec. \ref{sec:offaxis}. In Figs. \ref{fig:LIGO_HE} and \ref{fig:ET_HE}, we visualise the redshift distribution of our predicted GW and sGRB observations, for LIGO (design sensitivity) and ET respectively. These figures do not only generally provide insights into joint detection rates, but also give information on sub-threshold and off-axis joint observations. Figs. \ref{fig:LIGO_HE}-\ref{fig:halfET_HE} all use the same \textit{Fermi}/GBM GRB distribution. In this Section, we will discuss the former topic of these three, whereas he latter two will be addressed in Sec. \ref{sec:sub} and \ref{sec:offaxis}.

A feat that immediately stands out in Tab. \ref{tab:GW_sub} is the substantial difference in yearly joint detections between LIGO and ET. This is visualised in Figs. \ref{fig:LIGO_HE} and \ref{fig:ET_HE}. To accompany these, in Fig. \ref{fig:LIGOO3_HE} we also show the redshift distribution of \textit{Fermi}/GBM with LIGO-O3. In these figures, joint detections can only be made in the overlapping region. We see that the LIGO horizon is a main cause of the small number of joint GW-EM observations in our current era. Typical observable sGRBs by \textit{Fermi}/GBM lie at redshifts $\sim$0.6-0.8, while even at design sensitivity LIGO only reaches up to redshift $\sim$0.1. For LIGO-O3, the overlap is very minimal and hardly visible. For this configuration, we expect $\sim$0.027 joint detections per year out of $\sim$0.681 GW detections per year. This is a surprisingly low number of joint observations, given that we in fact had one joint detection in O2 \citep{abbott_multi-messenger_2017}. We refer to Sec. \ref{sec:offaxis} for an explanation. Even though LIGO-design shows an increase in joint detections with \textit{Fermi}/GBM ($\sim$0.141 yr$^{-1}$) over LIGO-O3, drastic instrument changes would have to be made in order to significantly extend the horizon out to where more BNSs can be observed gravitationally.

Our predictions for ET show the effects of such changes, depicted in Fig. \ref{fig:ET_HE}. Not only does ET observe significantly more sources than LIGO up to a redshift of 0.1, its horizon stretches out to much larger redshifts, clearly increasing the number of joint observations. We find that ET makes sure that the GW coverage for each HE detector in our model is close to 100\%, i.e. almost every observed GRB will also have a detected GW counterpart, as the joint observation rates found in Tab. \ref{tab:GW_sub} are very close to the yearly detection rates by the HE instruments themselves. This is an important finding, as it will allow us to place constraints on sGRB progenitors: if all sGRBs are accompanied by a GW signal, we will know that compact binaries are the sole origins of these bursts. It is important to realise that this is a theoretical prediction, as the GW detectors do not have a 100\% duty cycle, so in reality the GW coverage for every HE instument will be lower. Additionally, it should be noted that the while HE detectors that we use in the \GWT will not be operational during the ET era, we can still use them as reference instruments to extract useful information, some of which will apply to future instruments as well. With future HE detectors such as AMEGO \citep{mcenery_all-sky_2019}, HERMES \citep{fiore_hermes-technologic_2020}, TAP-GTM \citep{camp_transient_2019} and THESEUS \citep{amati_theseus_2018,amati_theseus_2021, ciolfi_multi-messenger_2021, rosati_synergies_2021}, we will be able to make joint observations at redshifts far beyond the current \textit{Fermi}/GBM limit of $\sim$2.


For illustrative purposes, we also show a "half-ET" configuration in of Fig. \ref{fig:halfET_HE}. This detector has a sensitivity that is halfway in-between LIGO (design sensitivity) and ET, which we obtained by multiplying the ET noise curve with a factor 10.89 (see \cite{yi_gravitational_2022}), representing an early stage of the ET capability or GW detectors in-between LIGO-design and ET-design. This configuration yields $\sim$408 GW observations per year, $\sim$6 of which would be observed by \textit{Fermi}/GBM as well. This shows that even an ET operating at a fraction of its capacity would be a major improvement over current-era detectors. As it is expected that ET will not immediately operate at its designed sensitivity, this is a very bright prospect. This is especially the case for the number of joint detections, as the "half-ET" horizon lies at a redshift of $\sim$1, which is much further than the typical LIGO horizon, leading to more and higher-redshift joint detections.

%A main factor influencing the number of joint detections is the redshift distribution of the BNS population observed by the corresponding GW and HE detector. In Fig. XX we plot in blue the z-distribution of BNSs observed by LIGO (design sensitivity), and in red the distribution of observable GRBs by \textit{Fermi}/GBM according to the Toolbox. We see that typical observable sGRBs by \textit{Fermi}/GBM lie at redshifts $\sim$0.6-0.8, while LIGO only reaches up to redshift $\sim$0.1. As such, the joint LIGO-Fermi detection fraction can never be high.

%Briefly compare HE detectors: why does Swift have so few, and why does GECAM have so many more for ET?

\subsection{Sub-threshold detections}
\label{sec:sub}
% Please add the following required packages to your document preamble:
% \usepackage{graphicx}
% Please add the following required packages to your document preamble:
% \usepackage{graphicx}
% Please add the following required packages to your document preamble:
% \usepackage{graphicx}
\begin{table*}[]
\centering
\renewcommand{\arraystretch}{1.5}
\resizebox{.75\textwidth}{!}{%
\begin{tabular}{c|cccc|cccc}
\multicolumn{9}{c}{Expected joint detection rates for different SNR thresholds} \\
\hline
\textbf{SNR} & \multicolumn{4}{c|}{\textbf{LIGO}} & \multicolumn{4}{c}{\textbf{ET}} \\ \hline
 & \textbf{GW} & \textbf{\textit{Fermi}/GBM} & \textbf{\textit{Swift}/BAT} & \textbf{GECAM} & \textbf{GW} & \textbf{\textit{Fermi}/GBM} & \textbf{\textit{Swift}/BAT} & \textbf{GECAM} \\ \cline{2-9} 
8 & 3.330 & 0.141 & 0.0236 & 0.143 & 12470 & 48 & 7 & 134 \\
7 & 4.792 & 0.178 & 0.0258 & 0.182 & 15372 & 50 & 8 & 149 \\
6 & 7.033 & 0.252 & 0.0410 & 0.264 & 18951 & 52 & 8 & 155 
\end{tabular}%
}
\caption{Number of GW and joint detections per year, for three different GW detector SNRs. For more accurate results, we ran all simulations involving LIGO for 1000 years rather than a single year. The numbers are the average rates from running each simulation 50 and 25 times for combinations involving LIGO and ET respectively.}
\label{tab:GW_sub}
\end{table*}

% Please add the following required packages to your document preamble:
% \usepackage{graphicx}
\begin{table*}[]
\centering
\renewcommand{\arraystretch}{1.5}
\resizebox{.75\textwidth}{!}{%
\begin{tabular}{c|ccc|ccc}
\multicolumn{7}{c}{Expected joint detection rates for different sub-threshold values} \\
\hline
\textbf{Sub-threshold} & \multicolumn{3}{c|}{\textbf{LIGO}} & \multicolumn{3}{c}{\textbf{ET}} \\ \hline
 & \textbf{\textit{Fermi}/GBM} & \textbf{\textit{Swift}/BAT} & \textbf{GECAM} & \textbf{\textit{Fermi}/GBM} & \textbf{\textit{Swift}/BAT} & \textbf{GECAM} \\ \cline{2-7} 
1 & 0.141 & 0.0236 & 0.143 & 48 & 7 & 134 \\ %\\
0.5 & 0.143 & 0.0239 & 0.145 & 83 & 12 & 179 \\%&  \\
0.3 & 0.147 & 0.0248 & 0.154 & 130 & 21 & 227 \\%& 
\end{tabular}%
}
\caption{Number of joint detections with different sub-threshold values for the HE detectors. Also here, we ran all simulations involving LIGO for 1000 years, and the numbers are the average rates from running each simulation 50 and 25 times for combinations involving LIGO and ET respectively.}
\label{tab:HE_sub}
\end{table*}

In the multi-messenger astronomy era, we have the possibility of using GW observations to search for possible EM counterparts in data from HE instruments, as well as the other way around. Since this might lead to the discovery of sGRBs or GW signals that would otherwise not have been found, a quantitative investigation could provide us valuable insights. Here, we address the potential gains of a targeted search for sub-threshold observations of BNSs.

Tab. \ref{tab:GW_sub} depicts the number of GW and joint observations for different GW-HE combinations as predicted by the Toolbox, for SNR thresholds of 8, 7 and 6 respectively. With the \Tb, we predict that for a threshold of 6, about 35 of 100 detected events is of astrophysical origin\footnote{The detection purity is calculated as follows: $P_{\text{astron}}=\frac{N_\text{astro}}{N_{\text{astro}}+N_{\text{noise}}}$. $N_{\rm{astro}}$ is the number of detections from astrophysical sources, which is estimated with the \Tb; $N_{\rm{noise}}$ is that from noise, which is estimated using analytical formula of \cite{lynch_observational_2018}.}. With HE data at hand, it would be quite easy to filter out the real GW signals whose properties match with those of the detected GRBs. We find that the fraction of astrophysical events drops significantly beyond 6; at 5.5, we already find ourselves at about 1 out of 100 detections that has an astrophysical origin. However, we might be able to still pinpoint this single GW detection with sGRB data; it is generally up to the GW-EM community which lowest SNR value maximises the scientific output of such studies \cite{lynch_observational_2018}. Tab. \ref{tab:HE_sub} shows a type of results similar to Tab. \ref{tab:GW_sub}, but for sub-threshold sGRB detections. Here, we do not go down further than a fraction 0.3 of the current treshold, as the \Tb models the typical flux of the background as a fraction 0.2 of the threshold, i.e. we need to stay above the background.

Tab. \ref{tab:GW_sub} clearly shows that the LIGO sensitivity is an important limiting factor in the current era. Fig. \ref{fig:LIGO_HE} visualises this, as we observe that many of the sources between SNR=8 and SNR=6 are at higher redshift ($\geq$0.1, up to $\sim$0.15), beyond the LIGO horizon with SNR=8. In this case, decreasing the threshold shifts the LIGO horizon more towards the peak of the z-distribution of the HE detectors, thus increasing the number of joint detections. Decreasing the SNR to 6 would lead to an increase in joint detections of e.g. 179\% and 185\% for \textit{Fermi}/GBM and GECAM with LIGO, respectively. As such, targeted searches for such sub-threshold GW observations would be extremely beneficial not only for the number of joint detections, but also for the BNS merger history. As the weaker GW signals typically originate from those BNSs further away, this would allow us to investigate the BNS redshift distribution and merger history in more detail. Additionally, with a larger sample extending to higher redshifts, we would be able to test gravitational waveform models in a larger part of their parameter space. Interestingly, lowering the threshold for HE detectors does not yield substantially more joint observations, as we notice in Tab. \ref{tab:HE_sub}. While the number of sGRB detections increases for lower thresholds, the majority of them is beyond the LIGO horizon and therefore does not contribute to the joint detection rate. Only a few intrinsically faint sGRBs below redshift $\sim$0.1 are added to the number of joint detections. Therefore, we can conclude that searches for sub-threshold GRBs using LIGO detections should not be a priority. The yield of such campaigns would be very limited, and focus should be on sub-threshold GWs with information from sGRB data at hand instead.

Tabs. \ref{tab:GW_sub} and \ref{tab:HE_sub} show different patterns for the ET era. First of all, in Tab. \ref{tab:GW_sub} we see barely any increase in the number of joint detections as we lower the threshold and increase the number of GW observations by ET. This ties in well with our previous remarks on the fact that nearly 100\% of all GRBs in the ET era will be also detected gravitationally. The depletion of current-era HE detectors during the ET era that is clearly visible here can be interpreted as a motivation for the development of new HE detectors that can keep up with the ET sensitivity. Future detectors should cover at least part of redshifts $\sim$2-5 to make a significant contribution to the number of joint detections with ET. Still, without new detectors reaching to much larger redshifts, a search for sub-threshold observations in GRB data of instuments like the current ones during the ET era could yield many new joint detections. In Tab. \ref{tab:HE_sub}, we can see increases up to 270\% for \textit{Fermi}/GBM, and joint observation numbers as high as 227 per year for GECAM. Similarly to before, lowering the HE detection threshold leads to more GRBs with higher redshift, most of which have a GW counterpart. Sub-threshold searches should not only be focused on higher redshifts (beyond $\sim$2), but also on those BNSs with low inclination. Those GRBs that are further away are much more likely to be observable if their jet points in our line-of-sight; off-axis GRBs are too weak to be observed at such redshifts. BNSs with inclination angles below $\sim$15$^\circ$ should be prioritised, as this is the typical width of a jet opening angle \citep{sarin_linking_2022, fong_decade_2015,salafia_structure_2022}.

In Fig. \ref{fig:halfET_HE}, the shaded region around the main blue population visualises the BNS detections by the "half-ET" configuration that have a SNR between 6 and 8. This population contains $\sim$956 BNSs per year, which results in a total number of 1364 when added to the original 408 sources with SNR $\geq$8. Most sub-threshold events are contained at redshifts beyond 0.2-0.3, reaching out to $\ge$1. The total population of BNSs with an SNR of 6 or higher yields $\sim$21 joint detections with half-ET and \textit{Fermi}/GBM. This is clearly higher than the $\sim$6 joint detections for those with SNR $\geq$8. Where the redshift distribution of half-ET BNSs does not overlap completely with \textit{Fermi}/GBM sGRBs, including sub-threshold GW detections can partially take care of this limitation as Fig. \ref{fig:halfET_HE} shows us that many BNSs are added at the high-redshift end. This results in about 50\% of \textit{Fermi}/GBM detections that have a GW counterpart which is detectable by half-ET. Where this is clearly not close to the near 100\% predicted for ET, it shows that even a half-ET configuration can be extremely beneficial and valuable for the GW-EM community, especially when compared with current-era GW instruments.

\subsection{Off-axis detections}
\label{sec:offaxis}
The sole joint GW-GRB detection thus far contained an off-axis GRB \citep{abbott_multi-messenger_2017, goldstein_ordinary_2017, ioka_can_2018}. With the \Tb we can predict the expected detection rate of such GRBs. We report the following results: we find $\sim$0.0035 joint observations with LIGO-O3 and \textit{Fermi}/GBM per year that are off-axis (with a general joint detection rate of $\sim$0.027 yr$^{-1}$). Assuming this number is realistic, chances that the very first joint GW-sGRB observation contains an off-axis GRB are very slim, making GW170817/GRB170817A an extremely unlikely event. A more logical explanation is the fact that these numbers expose the limitations of the top-hat jet model. In reality, the energy of the GRB leaks to a wider angle than we describe in this model, leading to observable (off-axis) GRBs whose inclination is well outside the jet cone. As such, with our current model, we treat many GRBs with higher inclination angles as undetected, where in reality they are visible to HE instruments.
%\g{in 3.3. you say 0.0254}

In this Section, we improve our off-axis detection rate by adding a second component to the model. Typically, a jet is surrounded by a wide-angle, mildly relativistic cocoon surrounding the relativistic jet \citep{zhang_physics_2018, ciolfi_short_2018}. We model this cocoon as the main source of off-axis detections. Still, since the vast majority of independent HE detections of sGRBs consists of on-axis GRBs, the cocoon should be several orders of magnitude less energetic than the main jet, not significantly contributing to the HE instruments' detection rate. We model the wide-angle cocoon emission by running a second population of GRBs, next to our main population described in Tab. \ref{tab:BNS_hyper}. For this second population, we use $\log E_{\text{GRB, mean}} = 48$, $\gamma = 150$ and a typical $\Delta\theta_{\text{mean}} = 60 ^\circ$ with $\Delta\theta_{\text{std}} = 1 ^\circ$ (see e.g. \cite{ramirez-ruiz_events_2002}). All other parameters remain unchanged. With this population, we can still recover the properties of GRB170817A-like GRBs, this time without having to increase $\nu^\prime_0$ to atypical values.

We show our predicted off-axis detection rates in Tab. \ref{tab:offaxis}. This population satisfies the condition of not significantly increasing the number of independent HE detections as compared to the main population: we find, for example, 1-2 extra detections with \textit{Fermi}/GBM per year. The joint detection rates we obtain with LIGO are more in line with predictions made by previous studies (e.g. \cite{howell_joint_2019, clark_prospects_2015}). Interestingly, we find that for all HE detectors, the joint off-axis detection rate with LIGO exceeds the number on-axis GRB observations listed in Tab. \ref{tab:GW_sub}. Here, we can clearly see the impact of the inclination angle and redshift. With a typical cocoon opening angle about 4 times larger than the GRB jet opening angle, this second population generally contains more sources whose emission is in the line-of-sight of the detector. Moreover, even though their flux is relatively weak as compared to the main GRB jet itself, many of these GRBs are still detected due to their proximity. With the LIGO horizon at a redshift of only $\sim$0.1, a relatively high off-axis detection rate is not surprising. This shows that the properties and characteristics of GW170817/GRB170817A are not as unique as one might think at first glance; with this improvement included, an off-axis GRB as a first joint detection is in fact quite likely. For ET, the off-axis detections encompass only a small part of the total number of joint observations. When combined with the on-axis observations in Tab. \ref{tab:GW_sub}, we still find a total rate that is in agreement with predictions \citep{ronchini_perspectives_2022}. ET stretches out to a horizon (a redshift of $\sim$5) that is far beyond that of HE instruments observing off-axis GRBs (only $\sim$0.05 for \textit{Fermi}/GBM). As such, the majority of ET joint observations will contain on-axis GRBs. Still, we expect more low-redshift off-axis joint detections with ET as compared to LIGO, due to its higher sensitivity. This means that the ET era is a very bright prospect, also in this regard. A potential $\sim$20 joint off-axis detections per year with ET-GECAM and likely more with next-generation HE detectors will allow us to cover significant parts of the parameter space for off-axis GRBs. We will be able to understand the GRB emission mechanism in more detail by comparing with the high number of on-axis joint detections we will have. This may probe a higher level of understanding of GRB central engines.

% Please add the following required packages to your document preamble:
% \usepackage{graphicx}
\begin{table}[]
\centering
\renewcommand{\arraystretch}{1.5}
\begin{tabular}{l|ll}
\multicolumn{3}{c}{Expected off-axis joint detection rates} \\
\hline
\textbf{} & \textbf{LIGO} & \textbf{ET} \\ \hline
\textbf{\textit{Fermi}/GBM} & 0.435 & 1.88 \\
\textbf{\textit{Swift}/BAT} & 0.0630 & 0.200 \\
\textbf{GECAM} & 0.692 & 19.7 \\
%\textbf{Insight-HE} &  & 
\end{tabular}%
\caption{The expected detection rates for off-axis GRBs for different GW detector - HE instrument combinations. For LIGO, we ran the population for 1000 years to obtain a more accurate rate. The values are averages that we obtained by repeating the simulation 50 times for LIGO combinations, and 25 times for ET. }
\label{tab:offaxis}
\end{table}

\section{Summary \& future prospects}
\label{sec:summary}
Based on \GWT, we developed an algorithm that simulates the potential prompt GRB emission from BNS and BHNS mergers, and check their detectability against a list of known HE detectors. Our algorithm pipeline of sGRB detectability determination is plotted as flowcharts in Figs. \ref{fig:flow-ns} and \ref{fig:flow-nsbh}. We employed the algorithm to simulate observations of sGRBs and joint observations with GW and HE detectors of BNS and BHNS, and studied the prospect of sub-threshold observation strategies in both GW and HE observation.

In the list below, we summarise our main findings in this study. Afterwards, we will discuss several future prospects for additions to the model as well as possible topics to explore using the \GWT.
\begin{itemize}
    \item We have constructed a sGRB model for the \GWT using a top-hat jet, operating with an accuracy sufficient for the purposes of the \Tb. The addition of this model increases the number of possible uses for the \Tb making it a valuable tool to investigate the detectability and properties of GW sources in the multi-messenger astronomy era.
    \item Our findings stress the significance of the impact that ET will have on multi-messenger astronomy. Where currently the LIGO sensitivity is the main limiting factor to the relatively low number of observations, we can quantitatively see that ET completely turns this around and will detect BNSs at an unprecedented rate. We have showed that a "half-ET" instrument which is more than 10 times less sensitive than ET already increases the number of joint detections significantly.
    \item We find that in the ET era, the GW coverage will be close to 100\% for each HE detector. Almost every detected GRB will have a GW counterpart that is observed by ET. In reality, this percentage will be lower due to the fact that GW detectors do not have a 100\% duty cycle. Still, reaching 100\% coverage would allow us to confirm that compact binaries are the sole origin of sGRBs.
    \item A targeted search campaign in LIGO data for sub-threshold GW detections has the potential of being successful in increasing the number of joint observations. LIGO is not able to detect most independent GRB detections by HE instruments. With information about the source from GRB data at hand, it will be relatively simple to pinpoint GW signals within the LIGO datastream in terms of localisation, chirp time and parameter space. We show an increase in joint detections of e.g. $\sim$185\% for GECAM if we would be able to find GW detections down to an SNR of 6. This highlights the value of multi-messenger observations, as a sub-threshold search campaign will clearly increase the number of observed BNSs, as well as the parameter space in which these BNSs find themselves.
    \item In the ET era, this will be the other way around as we will have many more GW observations than GRB detections. Those GWs further away measured by ET can be used to search for sub-threshold GRBs. Since ET stretches quite far out, we will be able to find more (e.g. $\sim$170\% more if we search down to a threshold half of the current \textit{Fermi}/GBM threshold) joint detections if we perform a targeted search to sub-threshold GRBs. Such a search should be focused on high-redshift (e.g. beyond $\sim$2 for \textit{Fermi}/GBM) GRBs, as closeby BNSs will be covered by both HE and GW detectors. Additionally, such campaigns are more likely to be successful when focusing on BNSs with inclinations lower than $\sim15^\circ$, as this is the typical opening angle of a GRB jet.% Even though current HE instruments will not be operational at the same time as ET, our general predictions still hold for future instruments with increased sensitivity.
    \item The top-hat jet model in the \GWT generally predicts joint detection rates that are quite low compared to other predictions, especially with LIGO. This is a direct result of the low number of off-axis detections predicted by the model. We achieve more realistic rates by including a second population, modelling the wide-angle cocoon around the jet, showing that the majority of joint detections in the LIGO era originates from off-axis GRBs. This feat makes the characteristics of GRB170817A not as unlikely as one may think based on the top-hat model. With ET, $\sim$100\% of electromagnetically detected off-axis GRBs will have an GW counterpart. However, the ET horizon stretches far beyond the off-axis detection horizon of current HE instruments, so on-axis GRBs will dominate the ET joint detection rate.
    %\item Check 2204.01746 paper
\end{itemize}

Our model has a number of limitations that can be addressed in future work. Here, we list potential improvements that serve as future prospects for continuation of this work. Additionally, we address general applications of the newly upgraded \GWT now functioning as a multi-messenger simulation program.

%\section{Discussion}
%\label{sec:disc}
\subsection{Future improvements}
\begin{itemize}
    \item We employed the top-hat jet model in this study, and a nested wider and less energetic top-hat jet to mimic the off-axis emissions. With increasing evidences, people now believe the jet of GRB should be structured \citep{salafia_structure_2022}. Our methodology is compatible with a structured jet, by substituting the top-hat function $\Pi(\Omega)$ in the emissivity with a more general continuous angular distribution. The time-evolving spectrum in this case will not take the form in equation (\ref{eqn:FvT}), but should be integrated through equation (\ref{eq:flux_generic}) respectively. 
        %\blue{Structured jet: might understimate off-axis sGRB detection rate. Can be approximated with a second population with wider, less energetic jet cone.}
    \item The comoving spectrum indices $\alpha_B$, $\beta_B$ and $s$ are not sampled from a distribution like $\gamma$, $\nu^\prime_0$ etc., but share the values for all sGRBs in the catalogue. In reality these parameters can differ from burst to burst, depending on the local emission condition and mechanism. We argue that the influence on detectability of bursts from the diversity on these spectrum parameters is largely degenerate with that from $\gamma$  and $\nu^\prime_0$. Our current treatment is a trade-off between the generality and degrees-of-freedom of the model, although extension to include six more hyper-parameters to describing distributions of these spectrum parameters would be straightforward. Other possible improvements to the spectrum could be more along the lines of \cite{ioka_spectral_2019}, where we would implement an intrinsic angular dependence in both $\gamma$ and $\nu^\prime_0$.
    \item In our simulation, we assume every BNS merger or BHNS that have $R_{\text{t}}>R_{\text{ISCO}}$ is associated with prompt GRB emission. While there are studies show that some merger remnants can not work as engine to power a GRB jet \citep{sarin_evolution_2020,shemi_appearance_1990,ciolfi_short_2018,nakar_short-hard_2007,murguia-berthier_necessary_2014,murguia-berthier_properties_2017,margalit_does_2015,beniamini_constraints_2017}, and the scenario of a "choked jet" is also possible; here, the jet fails to break out of the surrounding ejecta, and is not able to dissipate and emit a prompt GRB \citep{rosswog_high-resolution_2002,murguia-berthier_necessary_2014,lazzati_late_2018,bromberg_propagation_2011,lazzati_jet-cocoon_2019,salafia_gamma-ray_2020}. The above mentioned possibility will yield a fraction of failed GRB, which is not considered in our treatment, and will cause an overestimation of the sGRB detection rate by a factor $f_{\text{fail}}$; On the other hand, there could be sGRB whose progenitor is other than BNS/BHNS merger, which we also did not include in our consideration, which will cause an underestimation of the detection rate by a factor $f_{\text{other}}$. Since we collaborate the simulator with the historical detection rates, those effect of $f_{\text{other}}f_{\text{fail}}$ is absorbed into the merger rate parameter $R_{\text{n}}$ and the detector correcting factor $f_c$, and therefore will not cause under- or overestimation on the detection rate of future (joint) observations. When $R_{\text{n}}$ is more accurately estimated from a larger GW population, and $f_c$ is determined through dedicated trigger simulations, $f_{\text{fail}}f_{\text{other}}$ will be in turn constrained. 
\end{itemize}
\subsection{Applications on GRB physics and compact binaries merger history}
As shown inTab. \ref{tab:BNS_hyper}, the underlying GRB population model of the simulator has hyper-parameters that describe the merger rate of the progenitor binaries and GRB physics. A comparison between the simulated catalogue with the real observation from the same detector can in turn give constraints on this parameters, and thus give information on the GRB physics and progenitor binaries merger history. More specifically, such inference can be done in a Bayesian fashion: 
\begin{equation}
    p(\mathcal{B}|\{\Theta_i\})\propto\mathcal{L}(\{\Theta_i\}|\mathcal{B})p(\mathcal{B}),
\end{equation}
%-----------------------------------------------------------------
where $p(\mathcal{B}|\{\Theta_i\})$ is the posterior distributions of the list of the hyper-parameters of GRB population model, given a catalogue of detected GRB $\{\Theta_i\}$. $\Theta_i$ denotes a list of observable, e.g.,T90, Fluence, redshift, $E_{\rm{peak}}$ \textit{etc.}, of the $i$-th sGRB in the catalogues.

The key step is to find out the likelihood function on the right-hand side: $\mathcal{L}(\{\Theta_i\}|\mathcal{B})$, which has a known function form as in a inhomogeneous Poisson process:
\begin{equation}
    \mathcal{L}(\{\Theta_i\}|\mathcal{B})=N!\prod_{i=1}^Np(\Theta_{i,\rm{obs}}|\mathcal{B})p(N|\lambda_{\rm{GRB}}(\mathcal{B})),
\end{equation}
where $N$ is the number of event in the catalogue, $z_i$ is the redshift of $i$-th sGRB, $p(z_{i,\rm{obs}}|\mathcal{B})$ is probability of observing $z_i$ given model, and $\lambda_{\rm{GRB}}$ is the expected total number in the catalogue, which is also function of model parameters $\mathcal{B}$. $p(N|\lambda_{GRB})$ is a Poisson probability with expectation $\lambda_{\rm{GRB}}$. With simulations from \GWT, $\lambda_{GRB}$ and $p(\Theta_{i,\rm{obs}}|\mathcal{B})$ can be obtained, and thus makes the study possible. %\blue{Shuxu: This is a work that we could do after this paper}

% WARNING
%-------------------------------------------------------------------
% Please note that we have included the references to the file aa.dem in
% order to compile it, but we ask you to:
%
% - use BibTeX with the regular commands:
%   \bibliographystyle{aa} % style aa.bst
%   \bibliography{Yourfile} % your references Yourfile.bib
%
% - join the .bib files when you upload your source files
%-------------------------------------------------------------------

\bibliographystyle{aa}
\bibliography{aanda.bib}
\end{document}